\documentclass[12pt]{iopart}

\usepackage{graphicx}
\usepackage{dcolumn}
\usepackage{subfigure}
\usepackage{url}
\usepackage{bm}
\usepackage{color}
\usepackage{colortbl}

\begin{document}

\title[]{Musical Genres: Beating to the Rhythms of Different Drums}

\author{Debora C. Correa$^1$, Jose H. Saito$^2$ and Luciano da F. Costa$^1$}
\address{$^1$ Institute of Physics of Sao Carlos - University of Sao Paulo, Avenue
Trabalhador Sao Carlense 400, Caixa Postal 369, CEP 13560-970, Sao Carlos,
Sao Paulo, Brazil}

\address{$^2$ Computer Departament - Federal University of Sao Carlos, Rodovia Washington Luis, km 235,SP-310, CEP 13565-905, Sao Carlos,
Sao Paulo, Brazil}

\address{$^3$ National Institute of Science and Technology for Complex Systems, Brazil}
\eads{\mailto{deboracorrea@ursa.ifsc.usp.br}, \mailto{luciano.if.sc.usp.br}}

\begin{abstract}
Online music databases have increased signicantly as a consequence of
the rapid growth of the Internet and digital audio, requiring the
development of faster and more efficient tools for music content
analysis. Musical genres are widely used to organize music
collections. In this paper, the problem of automatic music genre
classification is addressed by exploring rhythm-based features
obtained from a respective complex network representation. A Markov
model is build in order to analyse the temporal sequence of rhythmic
notation events. Feature analysis is performed by using two
multivariate statistical approaches: principal component analysis
(unsupervised) and linear discriminant analysis
(supervised). Similarly, two classifiers are applied in order to
identify the category of rhythms: parametric Bayesian classifier under
gaussian hypothesis (supervised), and agglomerative hierarchical
clustering (unsupervised). Qualitative results obtained by Kappa
coefficient and the obtained clusters corroborated the effectiveness
of the proposed method.

\end{abstract}

\maketitle

\section{Introduction} \label{sec:Int}

Musical databases have increased in number and size continuously,
paving the way to large amounts of online music data, including
discographies, biographies and lyrics.  This happened mainly as a
consequence of musical publishing being absorbed by the Internet, as
well as the restoration of existing analog archives and advancements
of web technologies.  As a consequence, more and more reliable and
faster tools for music content analysis, retrieval and description,
are required, catering for browsing, interactive access and music
content-based queries. Even more promising, these tools, together with
the respective online music databases, have open new perspectives to
basic investigations in the field of music.

Within this context, music genres provide particularly meaningful
descriptors given that they have been extensively used for years to
organize music collections. When a musical piece becomes associated to
a genre, users can retrieve what they are searching in a much faster
manner. It is interesting to notice that these new possibilities of
research in music can complement what is known about the trajectories
of music genres, their history and their dynamics \cite{LENA2008}. In
an ethnographic manner, music genres are also particularly important
because they express the general identity of the cultural foundations
in which they are comprised \cite{Holt2007}. Music genres are part of
a complex interplay of cultures, artists and market strategies to
define associations between musicians and their works, making the
organization of music collections easier
\cite{SCARINGELLA2006}. Therefore, musical genres are of great interest 
because they can summarise some shared characteristics in music
pieces. As indicated by~\cite{AUCOUNTURIER2005}, music genre is
probably the most common description of music content, and its
classification represents an appealing topic in Music Information
Retrieval (MIR) research.

Despite their ample use, music genres are not a clearly defined
concept, and their boundaries remain fuzzy
\cite{SCARINGELLA2006}. As a consequence, the development of such
taxonomy is controversial and redundant, representing a challenging
problem. Pachet and Cazaly \cite{PACHET2000} demonstrated that there
is no general agreement on musical genre taxonomies, which can depend
on cultural references. Even widely used terms such as
\textit{rock}, \textit{jazz}, \textit{blues} and \textit{pop} are not
clear and firmly defined. According to~\cite{SCARINGELLA2006}, it is
necessary to keep in mind what kind of music item is being analysed in
genre classification: a song, an album, or an artist. While the most
natural choice would be a song, it is sometimes questionable to
classify one song into only one genre. Depending on the
characteristics, a song can be classified into various genres. This
happens more intensively with albums and artists, since nowadays the
albums contain heterogeneous material and the majority of artists tend
to cover an ample range of genres during their careers. Therefore, it
is difficult to associate an album or an artist with a specific
genre. Pachet and Cazaly~\cite{PACHET2000} also mention that the
semantic confusion existing in the taxonomies can cause redundancies
that probably will not be confused by human users, but may hardly be
dealt with by automatic systems, so that automatic analysis of the
musical databases becomes essential. However, all these critical
issues emphasize that the problem of automatic classification of
musical genres is a nontrivial task. As a result, only local
conclusions about genre taxonomy are considered
\cite{PACHET2000}.

As other problems involving pattern recognition, the process of
automatic classification of musical genres can usually be divided into
the following three main steps: representation, feature extraction and
the classifier design~\cite{COSTA2001,DUDA2001}. Music
information can be described by symbolic representation or based on
acoustic signals \cite{CATALTEPE2007}. The former is a high-level kind
of representation through music scores, such as MIDI, where each note
is described in terms of pitch, duration, start time and end time, and
strength. Acoustic signals representation is obtained by sampling the
sound waveform. Once the audio signals are represented in the
computer, the objective becomes to extract relevant features in order
to improve the classification accuracy. In the case of music, features
may belong to the main dimensions of it including melody, timbre,
rhythm and harmony. 

After extracting significant features, any classification scheme may
be used. There are many previous works concerning automatic genre classification
in the literature \cite{LI2009, MOSTAFA2009, WANG2008, SONG2008, PANAGAKIS2008, 
HONG2008, HOLZAPFEL2007, CATALTEPE2007, SILLA2007, SCARINGELLA2005,
SHAO2004, BURRED2003}.

An innovative approach to automatic genre classification is proposed
in the current work, in which musical features are referred to the
temporal aspects of the songs: the rhythm. Thus, we propose to
identify the genres in terms of their rhythmic patterns. While there
is no clear definition of rhythm \cite{SCARINGELLA2006}, it is
possible to relate it with the idea of temporal regularity. More
generally speaking, rhythm can be simply understood as a specific
pattern produced by notes differing in duration, pause and
stress. Hence, it is simpler to obtain and manipulate rhythm than the
whole melodic content. However, despite its simplicity, the rhythm is
genuine and intuitively characteristic and intrinsic to musical
genres, since, for example, it can be use to distinguish between rock
music and rhythmically more complex music, such as salsa. In addition,
the rhythm is largely independent on the instrumentation and
interpretation.

A few related works that use rhythm as features in automatic genre
recognition can be found in the literature. The work of Akhtaruzzaman
\cite{AKHTARUZZAMAN2008}, in which rhythm is analysed in
terms of mathematical and geometrical properties, and then fed to a
system for classification of rhythms from different regions. Karydis
\cite{KARYDIS2006} proposed to classify intra-classical genres with
note pitch and duration features, obtained from their histograms. In
\cite{GOUYON2005, GOUYON2004}, a review of existing automatic
rhythm description systems are presented. The authors say that despite
the consensus on some rhtyhm concepts, there is not a single
representation of rhythm that would be applicable for different
applications, such as tempo and meter induction, beat tracking,
quantization of rhythm and so on. They also analysed the relevance of
these descriptors by mesuring their performance in genre
classification experiments.  It has been observed that many of these
approaches lack comprehensiveness because of the relatively limited
rhythm representations which have been adopted~\cite{SCARINGELLA2006}.

In the current study, an objective and systematic analysis of rhythm
is provided. The main motivation is to study similar and different
characteristics of rhythms in terms of the occurrence of sequences of
events obtained from rhythmic notations. First, the rhythm is
extracted from MIDI databases and represented as graphs or networks
\cite{ALBERT2002, NEWMAN2003}. More specifically, each type of note
(regarding their duration) is represented as a node, while the
sequence of notes define the links between the nodes.  Matrices of
probability transition are extracted from these graphs and used to
build a Markov model of the respective musical piece. Since they are
capable of systematically modeling the dynamics and dependencies
between elements and subelements \cite{BOOTH1967}, Markov models are
frequently used in temporal pattern recognition applications, such as
handwriting, speech, and music \cite{ROHRMEIER2006}. Supervised and an
unsupervised approaches are then applied which receive as input the
properties of the transition matrices and produce as output the most
likely genre. Supervised classification is performed with the Bayesian
classifier. For the unsupervised approach, a taxonomy of rhythms is
obtained through hierarchical clustering. The described methodology is
applied to four genres: blues, \textit{bossa nova}, reggae and rock,
which are well-known genres representing different tendencies.  A
series of interesting findings are reported, including the ability of
the proposed framework to correctly identify the musical genres of
specific musical pieces from the respective rhythmic information.

This paper is organized as follows: section
\ref{sec:MatAndMethods} describes the methodology, including the
classification methods; section \ref{sec:ResAndDisc} presents the
obtained results as well as their discussion, and section
\ref{sec:Conclusions} contains the concluding remarks and future
works.

\section{Materials and Methods}\label{sec:MatAndMethods}

Some basic conceps about complex networks as well as the proposed
methodology are presented in this section.

\subsection{Systems Representation by Complex Networks} \label{ssec:CompNetworks}
A complex network is a graph exhibiting intricate structure when
compared to regular and uniformly random structures.  There are four
main types of complex networks: weighted and unweighted digraphs and
weighted and unweighted graphs. The operations of simmetry and
thresholding can be used to transform a digraph into a graph and a
weighted graph (or weighted digraph) into an unweighted one,
respectively \cite{COSTA2007}.  A weighted digraph (or weighted direct
graph) $G$ can be defined by the following elements:

- Vertices (or nodes). Each vertex is represented by an integer number
$i=1,2,...,N$; $N(G)$ is the vertex set of digraph $G$ and $N$
indicates the total number of vertices $(\left|N(G)\right|)$.

- Edges (or links). Each edge has the form $(i,j)$ indicating a
connection from vertex $i$ to vertex $j$. The edge set of digraph $G$
is represented by $\epsilon(G)$, and $M$ is the total number of edges.

- The mapping $\omega:\epsilon(G)\mapsto R$, where $R$ is the set of
weight values.  Each edge $(i,j)$ has a weight $\omega(i,j)$ associated
to it. This mapping does not exist in unweighted digraphs.

Undirected graphs (weighted or unweighted) are characterized by the
fact that their edges have not orientation. Therefore, an edge $(i,j)$
in such a graph necessarily implies a connection from vertex $i$ to
vertex $j$ and from vertex $j$ to vertex $i$.  A weighted digraph can
be represented in terms of its weight matrices $W$. Each element of
$W$, $w_{ji} $, associates a weight to the connection from vertex $i$
to vertex $j$.  The table
\ref{tab:BasicConcepts} summarizes some fundamental concepts about
graphs and digraphs \cite{COSTA2007}.

\begin{table*}
    \centering
    \caption{Graphs and Digraphs Basic Concepts}
    \footnotesize
        \begin{tabular}{l|c|c}
        \hline
        & \textbf{Graphs} & \textbf{Digraphs}
        \\
        \hline\hline
        Adjacency & Two vertices $i$ e $j$ are &Concepts of \\
        &adjacent or neighbors if $a_{ij}  = 1$ &predecessors and successors. If \\
        &&$a_{ij}  \ne 0$, $i$ is predecessor of \\
        &&$j$ and $j$ is successor of $i$. \\
        &&Predecessors e successors  \\
        &&as adjacent vertices.\\
        \hline
        Neighborhood & Represented by $v(i)$, meaning  & Also represented by $v(i)$. \\
        &the set of vertices that are & \\	
        &neighbors to vertex $i$.&\\
        \hline
        Vertex degree & Represented by $k_i$, gives & There are two kinds of degrees: in-degree $k_i^{in} $\\
        &the number of connected edges & indicating the number of incoming edges; \\
        &to vertex $i$. It is computed as: & and out-degree $k_i^{out} $ \\
        &$k_i  = \sum\limits_j {a_{ij} }  = \sum\limits_j {a_{ji} } $ & indicating the number of outgoing edges: \\
        &&$k_i^{in}  = \sum\limits_j {a_{ji} } $   $k_i^{out}  = \sum\limits_j {a_{ij} } $ \\
        &&The total degree is defined as $k_i  = k_i^{in}  + k_i^{out} $ \\
        \hline
        Average degree & Average of $k_i$ considering all & It is the same for in- and out- degrees. \\
        &network vertices. &  \\
        &$\left\langle k \right\rangle  = \frac{1}{N}\sum\limits_i {k_i }  = \frac{1}{N}\sum\limits_{ij} {a_{ij} } $ & $\left\langle {k^{out} } \right\rangle  = \left\langle {k^{in} } \right\rangle  = \frac{1}{N}\sum\limits_{ij} {a_{ij} } $ \\
        \hline\hline
        \end{tabular}
        \label{tab:BasicConcepts}
\end{table*}

For weighted networks, a quantity called strength of vertex $i$ is
used to express the total sum of weights associated to each node. More
specifically, it corresponds to the sum of the weights of the
respective incoming edges ($s_i^{in} = \sum\limits_j {w_{ji} }$)
(in-strength) or outgoing edges ($s_i^{out} = \sum\limits_j {w_{ij}
}$) (out-strength) of vertex $i$.

Another interesting measurement of local connectivity is the
clustering coefficient. This feature reflects the cyclic structure of
networks, i.e.  if they have a tendency to form sets of densely
connected vertices. For digraphs, one way to calculate the clustering
coefficient is: let $m_i$ be the number of neighbors of vertex $i$ and
$l_i$ be the number of connections between the neighbors of vertex
$i$; the clustering coefficient is obtained as $cc(i) = l_i / m_i (m_i
- 1)$.
      
\subsection{Data Description} \label{ssec:DataDescription}

In this work, four music genres were selected: blues, \textit{bossa
nova}, reggae and rock. These genres are well-known and represent
distinct major tendencies. Music samples belonging to these genres are
available in many collections in the Internet, so it was possible to
select one hundred samples to represent each one of them.  These
samples were downloaded in MIDI format. This event-like format
contains instructions (such as notes, instruments, timbres, rhythms,
among others) which are used by a synthesizer during the creation of
new musical events \cite{MIRANDA2001}. The MIDI format can be
considered a digital musical score in which the instruments are
separated into voices.

In order to edit and analyse the MIDI scores, we applied the software
for music composition and notation called Sibelius
(\url{http://www.sibelius.com}). For each sample, the voice related to
the percussion was extracted. The percussion is inherently suitable to
express the rhythm of a piece. Once the rhythm is extracted, it
becomes possible to analyse all the involved elements. The MIDI
Toolbox for Matlab was used \cite{Eerola2004}. This Toolbox is free
and contains functions to analyse and visualize MIDI files in the
Matlab computing environment. When a MIDI file is read with this
toolbox, a matrix representation of note events is created. The
columns in this matrix refer to many types of information, such as:
onset (in beats), duration (in beats), MIDI channel, MIDI pitch,
velocity, onset (in seconds) and duration (in seconds). The rows refer
to the individual note events, that is, each note is described in
terms of its duration, pitch, and so on.

\begin{figure}
\centering
\subfigure[]
{
    \label{fig:exemplosredes:a}
    \includegraphics[width=7cm]{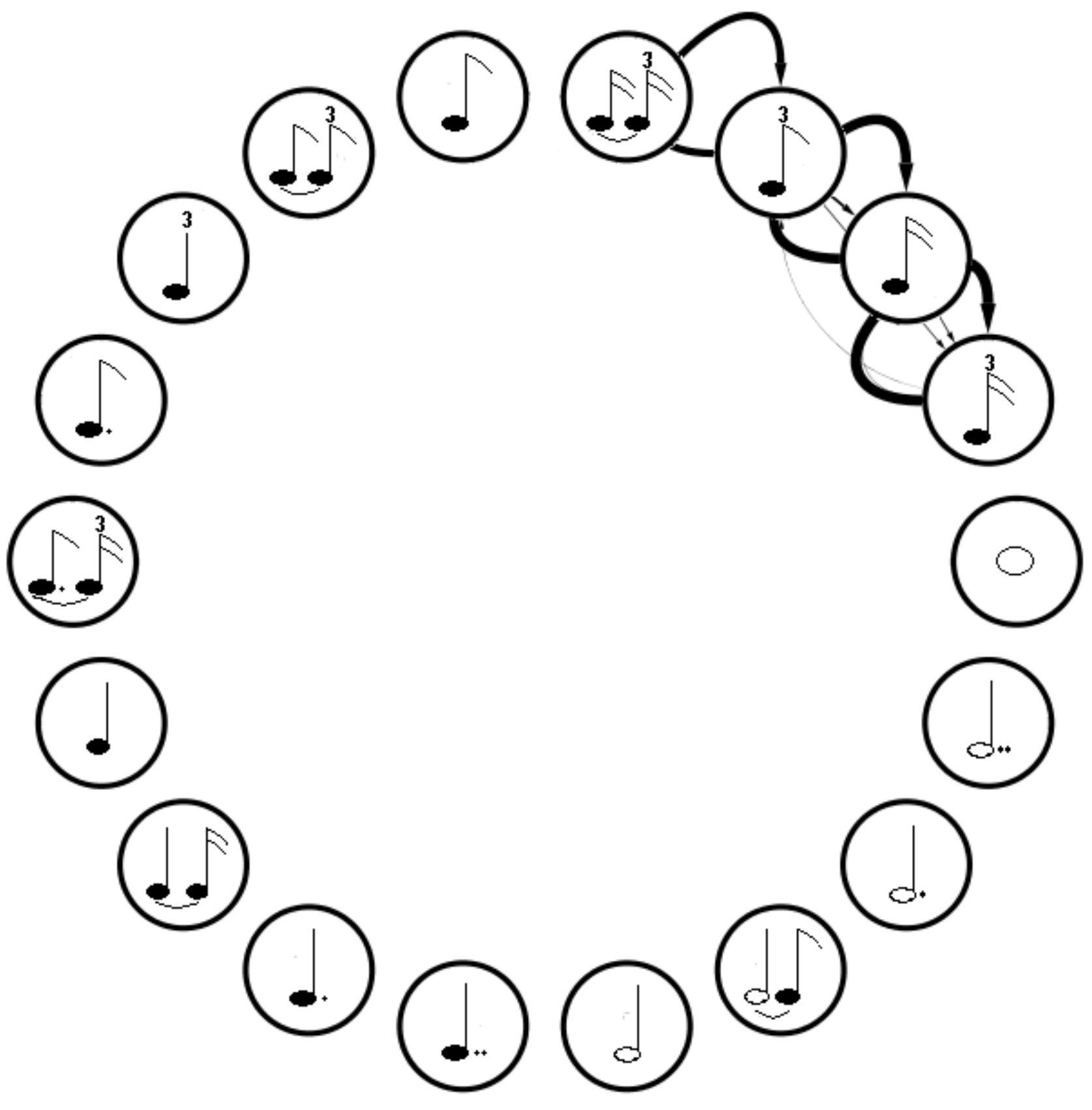}
}
\hspace{0.5cm}
\subfigure[]
{
    \label{fig:exemplosredes:b}
    \includegraphics[width=7cm]{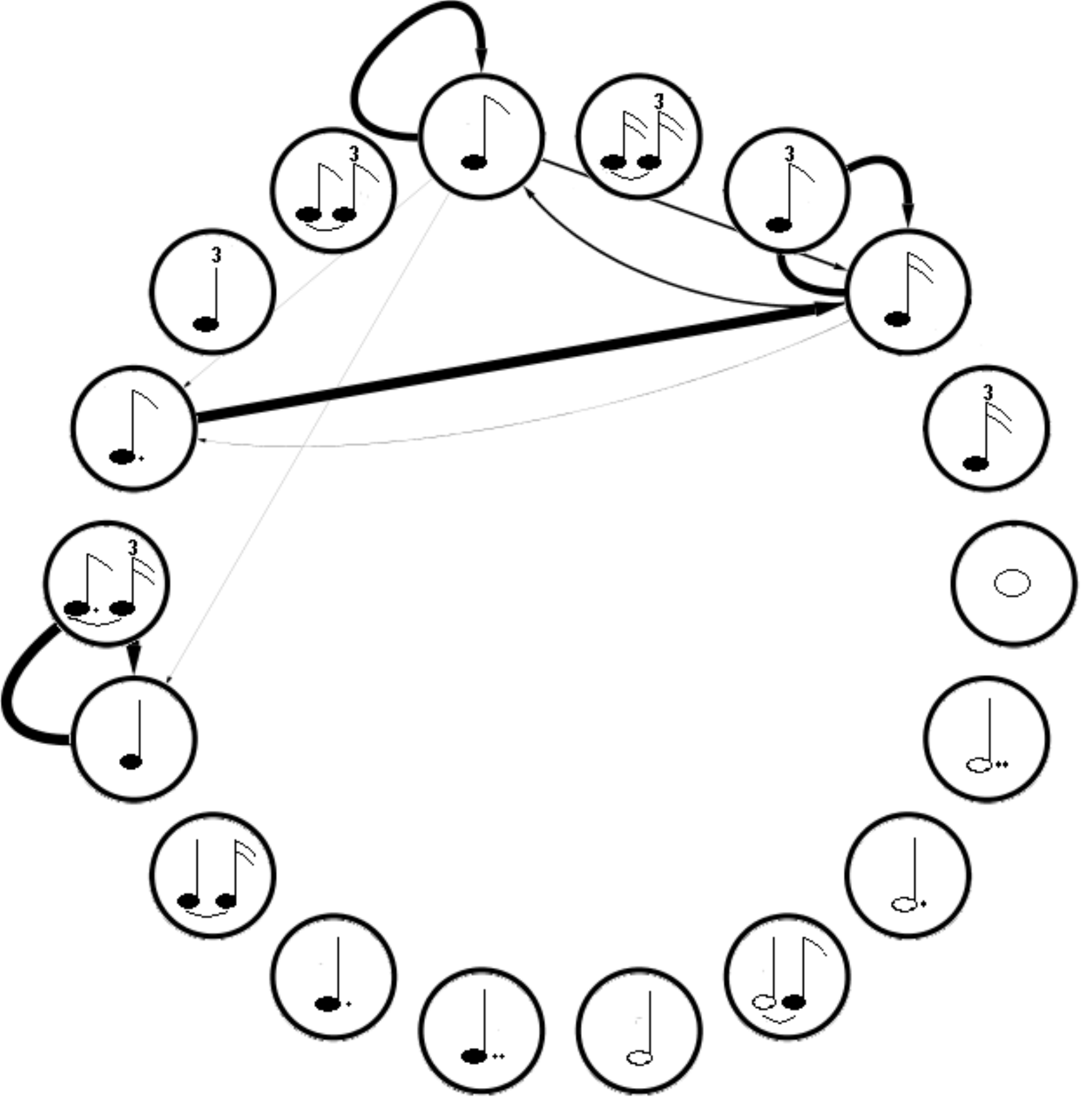}
}
\hspace{0.5cm}
\subfigure[]
{
    \label{fig:exemplosredes:c}
    \includegraphics[width=7cm]{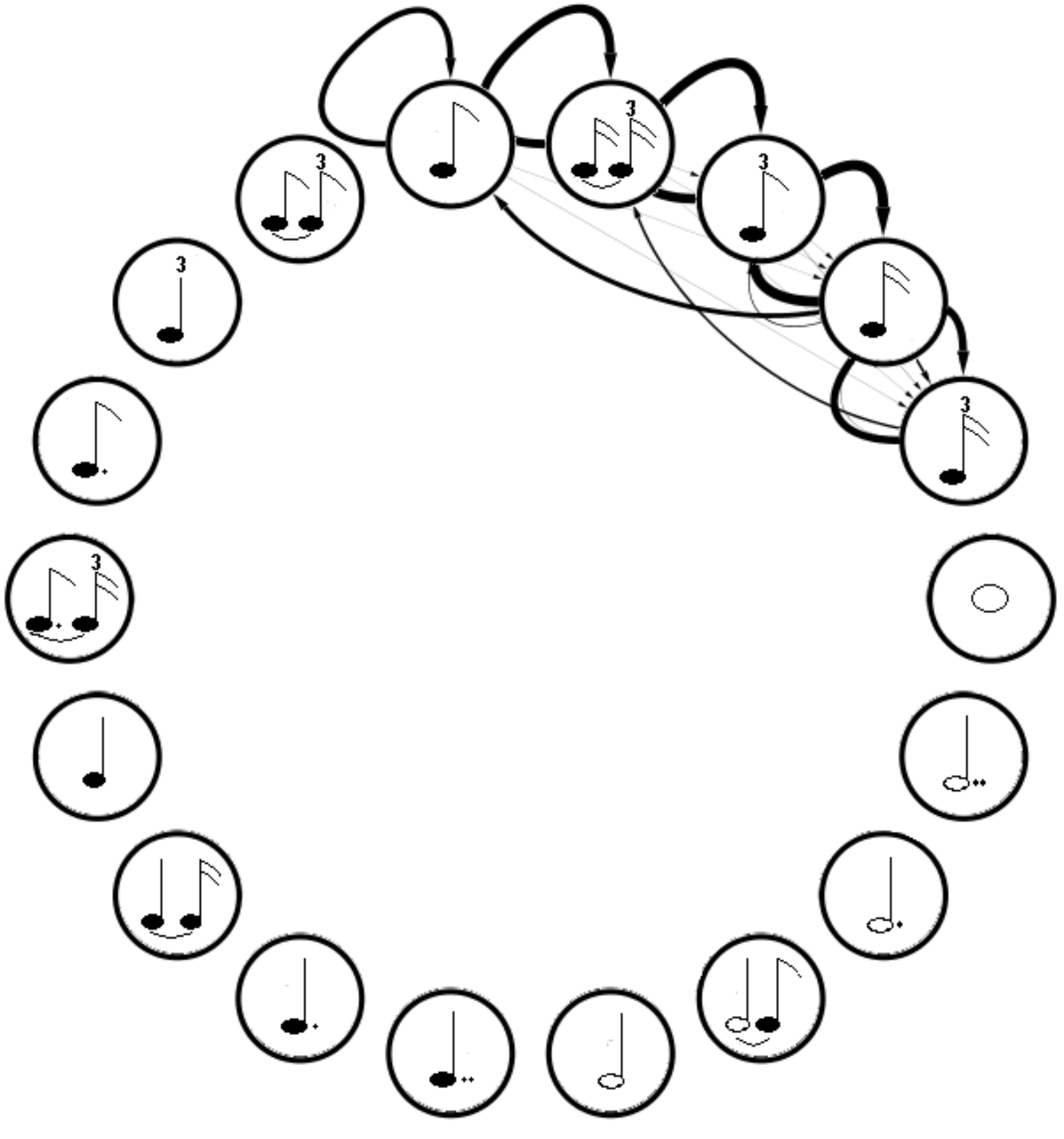}
}
\hspace{0.5cm}
\subfigure[]
{
    \label{fig:exemplosredes:d}
    \includegraphics[width=7cm]{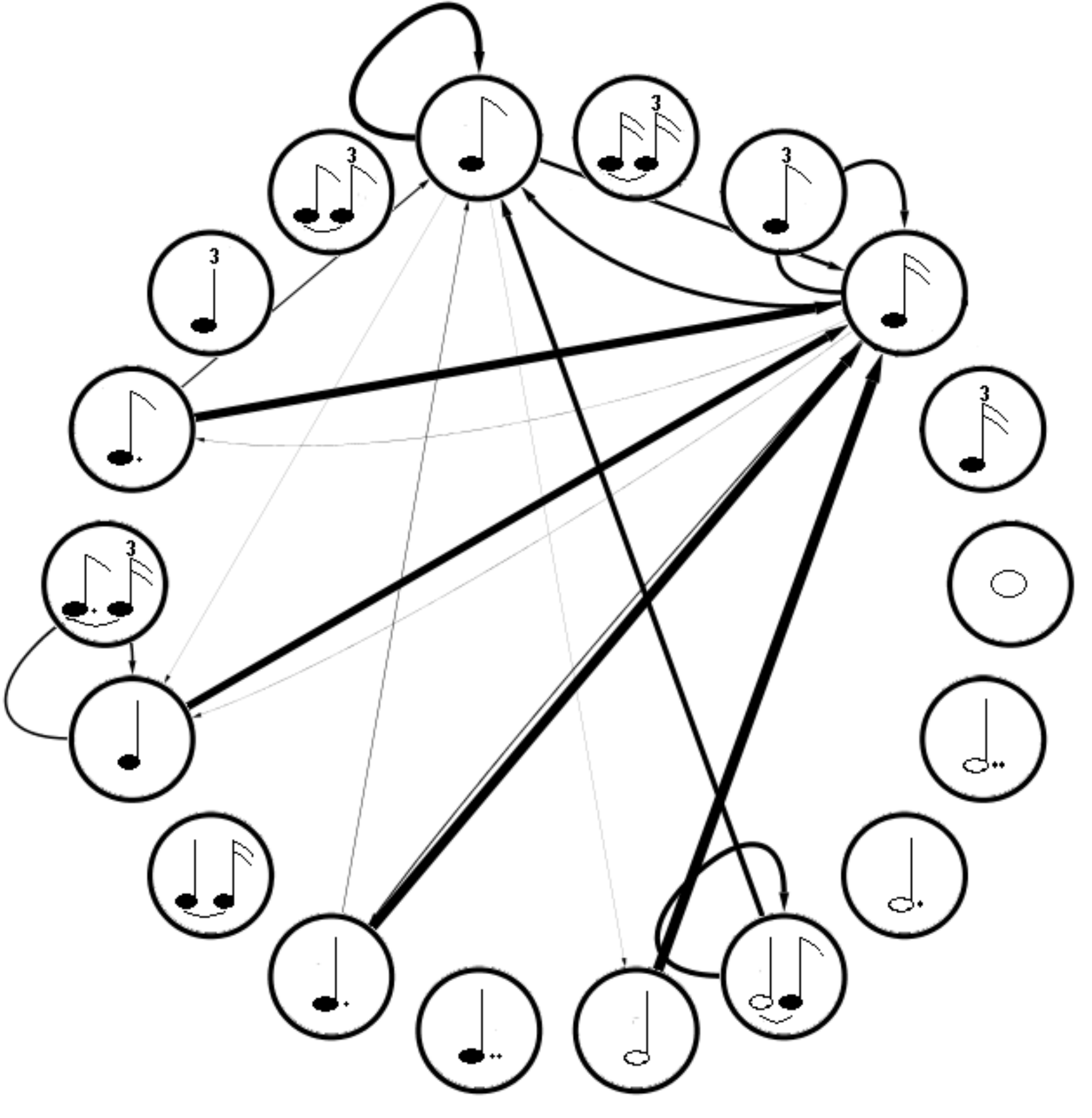}
}
  \caption{Digraph examples of four music samples: (a) \textit{How
  Blue Can You Get} by BB King. (b) \textit{Fotografia} by Tom
  Jobim.  (c) \textit{Is This Love} by Bob Marley. (d) \textit{From
  Me To You} by The Beatles.}
\label{fig:exemplosredes}

\end{figure}

Only the note duration (in beats) has been used in the current
work. In fact, the durations of the notes, respecting the sequence in
which they occur in the sample, are used to create a digraph. Each
vertex of this digraph represents one possible rhythm notation, such
as, quarter note, half note, eighth note, and so on. The edges reflect
the subsequent pairs of notes. For example, if there is an edge from
vertex $i$, represented by a quarter note, to a vertex $j$,
represented by an eighth note, this means that a quarter note was
followed by an eighth note at least once. The thicker the edges, the
larger is the strength between these two nodes.  Examples of these
digraphs are showed in Figure \ref{fig:exemplosredes}. Figure
\ref{fig:exemplosredes:a} depicts a blues sample represented by the
music \textit{How blue can you get} by BB King. A \textit{bossa nova} sample,
namely the music \textit{Fotografia} by Tom Jobim, is illustrated in
\ref{fig:exemplosredes:b}. Figure \ref{fig:exemplosredes:c} illustrates
a reggae sample, represented by the music \textit{Is this Love} by Bob
Marley. Finally, Figure \ref{fig:exemplosredes:d} shows a rock sample,
corresponding the music \textit{From Me To You} by The Beatles.

\subsection{Feature Extraction} \label{ssec:FeatureExtraction}

Extracting features is the first step of most pattern recognition
systems.  Each pattern is represented by its \textit{d} features or
attributes in terms of a vector in a \textit{d}- dimensional space. In
a discrimination problem, the goal is to choose features that allow
the pattern vectors belonging to different classes to occupy compact
and distinct regions in the feature space, maximizing class
separability.  After extracting significant features, any
classification scheme may be used.  In the case of music, features may
belong to the main dimensions of it including melody, timbre, rhythm
and harmony.

Therefore, one of the main features of this work is to extract
features from the digraphs and use them to analyse the complexities of
the rhythms, as well as to perform classification tasks. For each
sample, a digraph is created as described in the previous section. All
digraphs have 18 nodes, corresponding to the quantity of rhythm
notation possibilities concerning all the samples, after excluding
those that hardly ever happens. This exclusion was important in order
to provide an appropriate visual analysis and to better fit the
features. In fact, avoiding features that do not significantly
contribute to the analysis reduces data dimension, improves the
classification performance through a more stable representation and
removes redundant or irrelevant information (in this case, minimizes
the occurrence of null values in the data matrix).

The features are associated with the weight matrix $W$. As commented
in section \ref{ssec:CompNetworks}, each element in $W$, $w_{ij}$,
indicates the weight of the connection from vertex $j$ to $i$, or, in
other words, they are meant to represent how often the rhythm
notations follow one another in the sample. The weight matrix $W$ has
18 rows and 18 columns. The matrix $W$ is reshaped by a 1 x 324
feature vector. This is done for each one of the genre
samples. However, it was observed that some samples even belonging to
different genres generated exactly the same weight matrix. These
samples were excluded. Thereby, the feature matrix has 280 rows (all non-excluded
samples), and 324 columns (the attributes).

\begin{figure}
\centering
    \includegraphics[width=10cm]{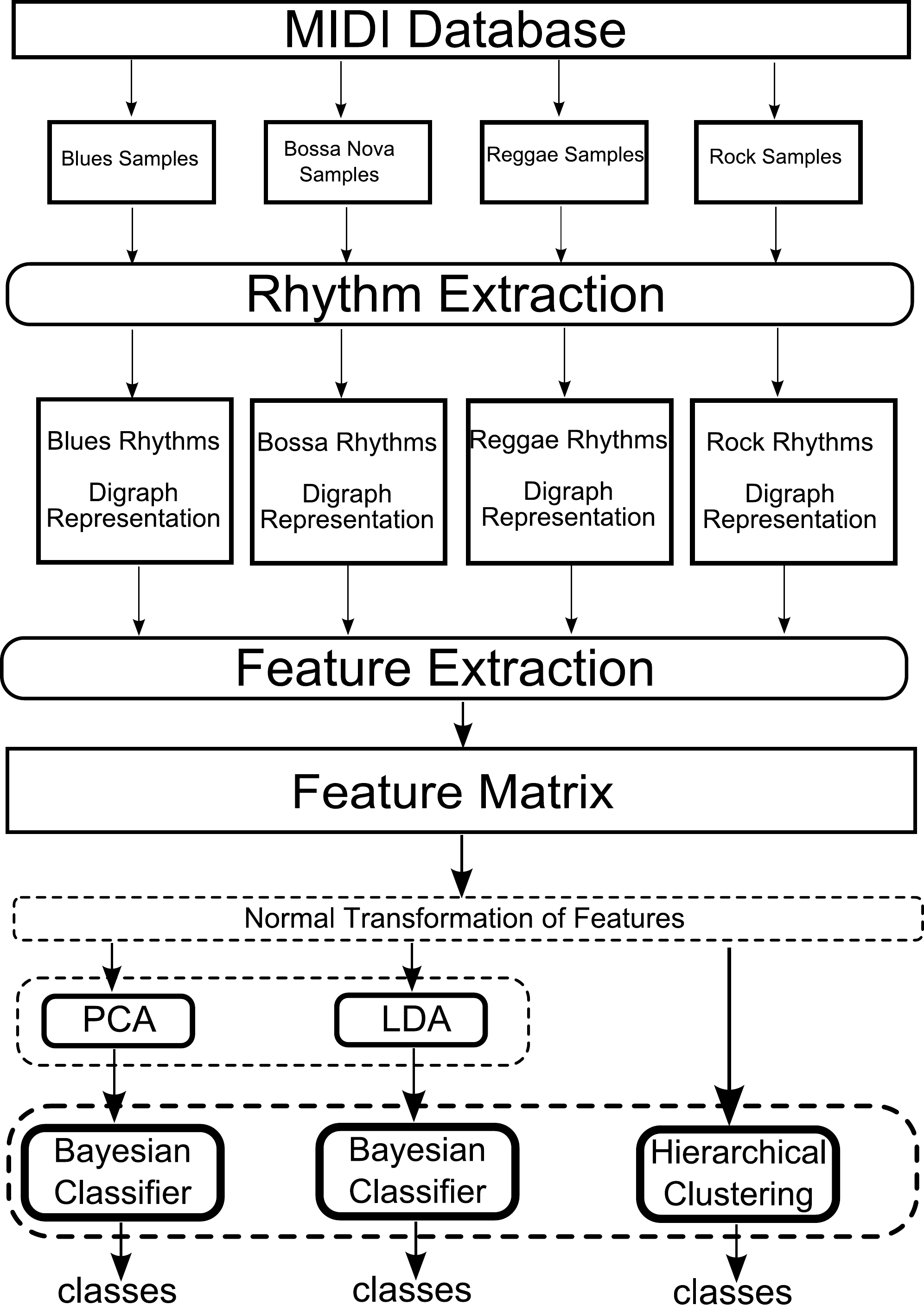}

\caption{Block diagram of the proposed methodology.} 
\label{fig:blockdiagram}
\end{figure}

An overview of the proposed methodology is illustrated in Figure
\ref{fig:blockdiagram}.  After extracting the features, a standardization
transformation is done to guarantee that the new feature set has zero
mean and unit standard deviation.  This procedure can significantly
improve the resulting classification.  Once the normalized features
are available, the structure of the extracted rhythms can be analysed
by using two different approaches for features analysis: PCA and LDA.
We also compare two types of classification methods: Bayesian
classifier (supervised) and hierarchical clustering
(unsupervised). PCA and LDA are described in section
\ref{subsec:FeatureAnalysis} and the classification methods are described in section \ref{subsec:ClassifMethodology}.

\subsection{Feature Analysis and Redundancy Removing} \label{subsec:FeatureAnalysis}

\begin{figure}
\centering
    \includegraphics[width=8cm]{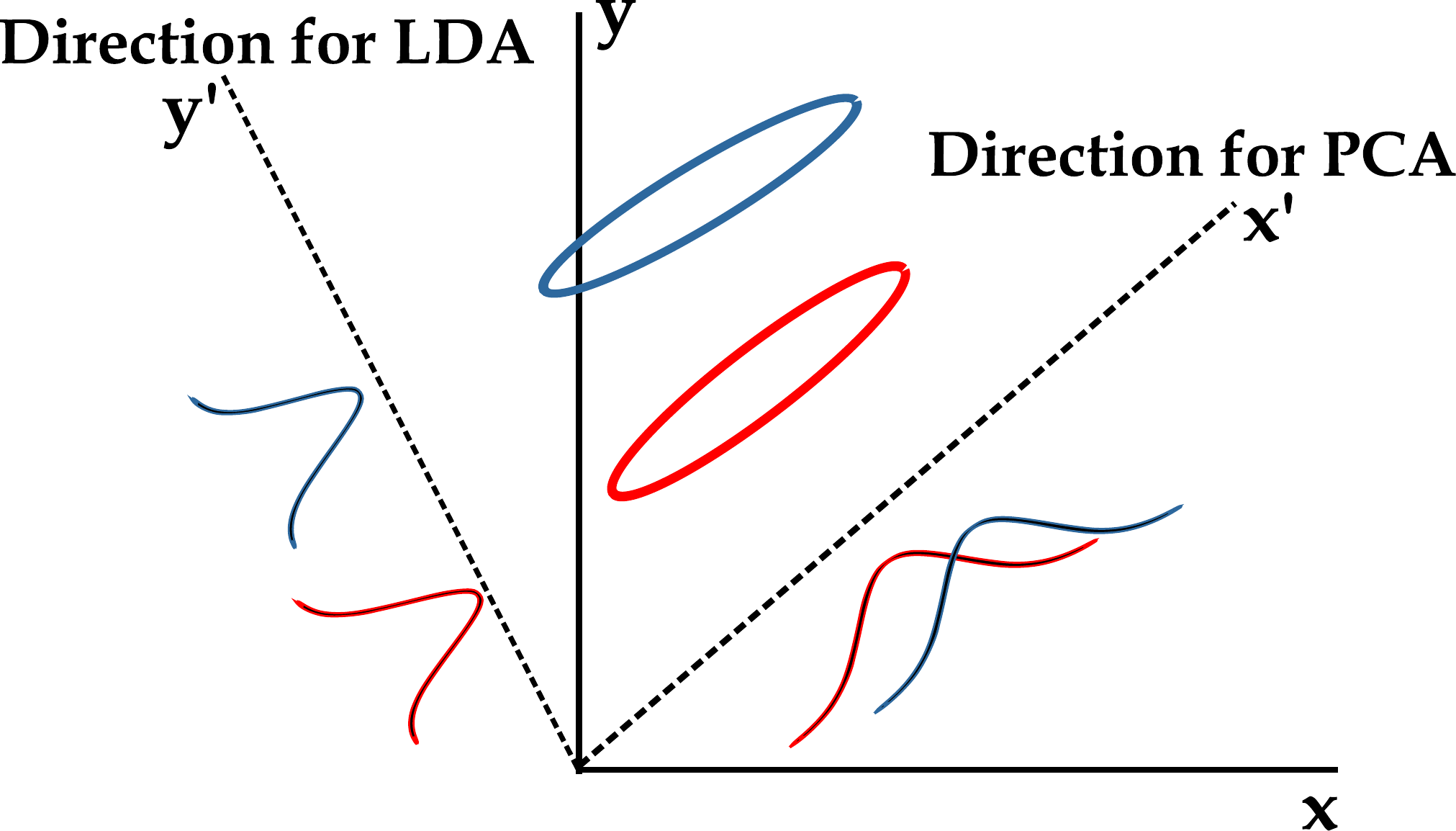}
  \caption{An illustration of PCA (optimizing
  representation) and LDA projections (optimizing classification),
  adapted from \cite{Therrien1989}.}
\label{fig:pcalda}
\end{figure}

Two techniques are widely used for feature analysis 
\cite{DEVIJVER1981, DUDA2001, WEBB2002}: PCA
(Principal Components Analysis) and LDA (Linear discriminant
Analysis).  Basically, these approaches apply geometric
transformations (rotations) to the feature space with the purpose of
generating new features based on linear combinations of the original
ones, aiming at dimensionality reduction (in the case of PCA) or to
seek a projection that best separates the data (in the case of
LDA). Figure\ref{fig:pcalda} illustrates the basic principles
underlying PCA and LDA. The direction \textbf{$x'$} (obtained with
PCA) is the best one to represent the two classes with maximum overall
dispersion.  However, tt can be observed that the densities projected
along direction \textbf{$x'$} overlap one another, making these two
classes inseparable. Differently, if direction \textbf{$y'$}, obtained
with LDA, is chosen, the classes can be easily separated. Therefore,
it is said that the directions for representation are not always also
the best choice for classification, reflecting the different
objectives of PCA and LDA. \cite{Therrien1989}. \ref{subsec:PCA} and \ref{subsec:LDA} give more details about these
two techniques.

\subsection{Classification Methodology} \label{subsec:ClassifMethodology}

Basically, to classify means to assign objects to classes or
categories according to the properties they present. In this context,
the objects are represented by attributes, or feature vectors. There
are three main types of pattern classification tasks: \textit{imposed
criteria}, \textit{supervised} classification and
\textit{unsupervised} classification. Imposed criteria is the easiest
situation in classification, once the classification criteria is
clearly defined, generally by a specific practical problem. If the
classes are known in advance, the classification is said to be
supervised or by example, since usually examples (the training set)
are available for each class. Generally, supervised classification
involves two stages: \textit{learning}, in which the features are
tested in the training set; and \textit{application}, when new
entities are presented to the trained system. There many approaches
involving supervised classification. The current study applied the
Bayesian classifier through discriminant functions
(\ref{sec:ALinearDiscFunc}) in order to perform supervised
classification. The Bayesian classifier is based on the Bayesian
Decision Theory and combines class conditional probability densities
(likelihood) and prior probabilities (prior knowledge) to perform
classification by assigning each object to the class with the maximum
a posteriori probability.

In unsupervised classification, the classes are not known in advance,
and there is not a training set. This type of classification is
usually called \textit{clustering}, in which the objects are
agglomerated according to some similarity criterion. The basic
principle is to form classes or \textit{clusters} so that the
similarity between the objects in each class is maximized and the
similarity between objects in different classes minimized. There are
two types of clustering: partitional (also called non-hierarchical)
and hierarchical. In the former, a fixed number of clusters is
obtained as a single partition of the feature space. Hierarchical
clustering procedures are more commonly used because they are usually
simpler \cite{DUDA2001, COSTA2001, WEBB2002, THEODORIDIS2006}.  The
main difference is that instead of one definite partition, a series of
partitions are taken, which is done progressively. If the hierarchical
clustering is \textit{agglomerative} (also known as bottom-up), the
procedure starts with $N$ objects as $N$ clusters and then
successively merges the clusters until all the objects are joined into
a single cluster (please refer to \ref{sec:HierClust} for more details
about agglomerative hierarchical clustering). \textit{Divisive}
hierarchical clustering (top-down) starts with all the objects as one
single cluster, and splits it into progressively finer subclusters. 
  
\subsubsection{Performance Measures for Classification} \label{subsec:PerfMeasClassif}

To objectively evaluate the performance of the supervised
classification it is necessary to use quantitative criteria. Most used
criteria are the estimated classification error and the obtained
accuracy. Because of its good statistical properties, as, for example,
be asymptotically normal with well defined expressions to estimate its
variances, this study also adopted the Cohen Kappa Coefficient
\cite{COHEN1960} as a quantitative measure to analyse the performance
of the proposed method. Besides, the kappa coefficient can be directly
obtained from the confusion matrix \cite{CONGALTON1991} (\ref{sec:AKappa}), easily
computed in supervised classification problems. The confusion matrix
is defined as:

\begin{equation}
\centering
C = \left[ {\begin{array}{*{20}c}
   {c_{11} } & {c_{12} } &  \ldots  & {c_{1c} }  \\
   {c_{21} } &  \ddots  & {} &  \vdots   \\
    \vdots  & {} &  \ddots  & {}  \\
   {c_{c1} } &  \ldots  & {} & {c_{cc} }  \\
\end{array}} \right]
\end{equation}

where each element $c_{ij} $ represents the number of objects from
class $i$ classified as class $j$. Therefore, the elements in the
diagonal indicate the number of correct classifications.

\clearpage
\section{Results and Discussion}\label{sec:ResAndDisc}

As mentioned before, four musical genres were used in this study:
blues, \textit{bossa nova}, reggae and rock. It was selected music art
works from diverse artists, as presented in Tables
\ref{tab:BluesAndBossa} and \ref{tab:ReggaeAndRock}. Different colors
were chosen to represent the genres (color red for genre blues, green
for \textit{bossa nova}, cyan for reggae and pink for rock), in order
to provide a better visualization and discussion of the results.

\begin{table}
\caption{Blues and \textit{Bossa nova} art works}
\begin{minipage}[b]{.40\textwidth}
      \begin{indented}
       \item[]
       \tiny
       \color{red}
       \arrayrulecolor{black}
        \begin{tabular}{l c c}
        \hline
        & \textcolor{black}{\textbf{Blues art works}}  \\
         \hline\hline
        1 &  Albert Collins- A  Good  Fooll\\ 
        2 &  Albert Collins- Fake  ID\\
				3 &  Albert King- Stormy  Monday\\
				4 &  Andrews Sister- Boogie  Woogie  Bugle  Boy\\
				5 &	 B B King- Dont  Answer  The  Door\\
				6 &  B B King- Get  Off  My  Back\\
				7 &  B B King- Good  Time\\
				8 &  B B King- How  Blue  Can  You  Get\\
				9 &  B B King- Sweet  Sixteen\\
				10 & B B King- The  Thrill  Is  Gone\\
				11 & B B King- Woke  Up  This  Morning\\
				12 & Barbra Streisand- Am  I  Blue\\
				13 & Billie Piper- Something  Deep  Inside\\
				14 & Blues In F For Monday\\
				15 & Bo Diddley- Bo Diddley\\
				16 & Boy Williamson- Dont Start Me Talking\\
				17 & Boy Williamson- Help Me\\
				18 & Boy Williamson- Keep It To Yourself\\
				19 & Buddy Guy- Midnight Train\\
				20 & Charlie Parker- Billies Bounce\\
				21 & Count Basie- Count On The Blues\\
				22 & Cream- Crossroads Blues\\
				23 & Delmore Brothers- Blues Stay Away From Me\\
				24 & Elmore James- Dust My Broom\\
				25 & Elves Presley- A Mess Of Blues\\
				26 & Etta James- At Last\\
				27 & Feeling The Blues\\
				28 & Freddie King- Help Day\\
				29 & Freddie King- Hide Away\\
				30 & Gary Moore- A Cold Day In Hell\\
				31 & George Thorogood- Bad To The Bone\\
				32 & Howlin Wolf- Little Red Rooster\\
				33 & Janis Joplin- Piece Of My Heart\\
				34 & Jimmie Cox- Before You Accuse Me\\
				35 & Jimmy Smith- Chicken Shack\\
				36 & John Lee Hooker- Boom Boom Boom\\
				37 & John Lee Hooker- Dimples\\
				38 & John Lee Hooker- One Bourbon One Scotch One Beer\\
				39 & Johnny Winter- Good Morning Little School Girl\\
				40 & Koko Taylor- Hey Bartender\\
				41 & Little Walter- Juke\\
				42 & Louis Jordan- Let Good Times Roll\\
				43 & Miles Davis- All Blues\\
				44 & Ray Charles- Born To The Blues\\
				45 & Ray Charles- Crying Times\\
				46 & Ray Charles- Georgia On My Mind\\
				47 & Ray Charles- Hit The Road Jack\\
				48 & Ray Charles- Unchain My Heart\\
				49 & Robert Johnson- Dust My Broom\\
				50 & Stevie Ray Vaughan- Cold Shot\\
				51 & Stevie Ray Vaughan- Couldnt Stand The Weather\\
				52 & Stevie Ray Vaughan- Dirty Pool\\
				53 & Stevie Ray Vaughan- Hillbillies From Outer Space\\
				54 & Stevie Ray Vaughan- I Am Crying\\
				55 & Stevie Ray Vaughan- Lenny\\
				56 & Stevie Ray Vaughan- Little Wing\\
				57 & Stevie Ray Vaughan- Looking Out The Window\\
				58 & Stevie Ray Vaughan- Love Struck Baby\\
				59 & Stevie Ray Vaughan- Manic Depression\\
				60 & Stevie Ray Vaughan- Scuttle Buttin\\
				61 & Stevie Ray Vaughan- Superstition\\
				62 & Stevie Ray Vaughan- Tell Me\\
				63 & Stevie Ray Vaughan- Voodoo Chile\\
				64 & Stevie Ray Vaughan- Wall Of Denial\\
				65 & T Bone Walker- Call It Stormy Monday\\
				66 & The Blues Brothers- Everybody Needs Somebody To Love\\
				67 & The Blues Brothers- Green Onions\\
				68 & The Blues Brothers- Peter Gunn Theme\\
				69 & The Blues Brothers- Soulman\\
				70 & W C Handy- Memphis Blues\\
        \hline\hline
\end{tabular}
\end{indented}
\end{minipage}\qquad
\begin{minipage}[b]{.40\textwidth}
    \begin{indented}
     \item[]
     \tiny
     \color[rgb]{0.2,0.8,0}
          \begin{tabular}{l c c}
        \hline
          & \textcolor{black}{\textbf{\textit{Bossa nova} art works}}  \\
        \hline\hline
        71 &   Antonio Adolfo- Sa Marina\\
				72 & Barquinho\\
				73 & Caetano Veloso- Menino Do Rio\\
				74 & Caetano Veloso- Sampa\\
				75 & Celso Fonseca- Ela E Carioca\\
				76 & Celso Fonseca- Slow Motion bossa nova\\
				77 & Chico Buarque Els Soares- Facamos\\
				78 & Chico Buarque Francis Hime- Meu Caro Amigo\\
				79 & Chico Buarque Quarteto Em Si- Roda Viva\\
				80 & Chico Buarque- As Vitrines\\
				81 & Chico Buarque- Construcao\\
				82 & Chico Buarque- Desalento\\
				83 & Chico Buarque- Homenagem Ao Malandro\\
				84 & Chico Buarque- Mulheres De Athenas\\
				85 & Chico Buarque- Ole O La\\
				86 & Chico Buarque- Sem Fantasia\\
				87 & Chico Buarque- Vai Levando\\
				88 & Dick Farney- Copacaba\\
				89 & Elis Regina- Alo Alo Marciano\\
				90 & Elis Regina- Como Nossos Pais\\
				91 & Elis Regina- Na Batucada Da Vida\\
				92 & Elis Regina- O Bebado E O Equilibrista\\
				93 & Elis Regina- Romaria\\
				94 & Elis Regina- Velho Arvoredo\\
				95 & Emilio Santiago- Essa Fase Do Amor\\
				96 & Emilio Santiago- Esta Tarde Vi Voller\\
				97 & Emilio Santiago- Saigon\\
				98 & Emilio Santiago- Ta Tudo Errado\\
				99 & Gal Costa- Canta Brasil\\
			 100 & Gal Costa- Para Machucar Meu Coracao\\
			 101 &	 Gal Costa- Pra Voce\\
			 102 &	 Gal Costa- Um Dia De Domingo\\
			 103 &	 Jair Rodrigues- Disparada\\
			104 &	 Joao Bosco- Corsario\\
			105 &	 Joao Bosco- De Frente Para O Crime\\
			106 &	 Joao Bosco- Jade\\
			107 &	 Joao Bosco- Risco De Giz\\
			108 &	 Joao Gilberto- Corcovado\\
			109 &	 Joao Gilberto- Da Cor Do Pecado\\
			110 &	 Joao Gilberto- Um Abraco No Bonfa\\
			111 &	 Luiz Bonfa- De Cigarro Em Cigarro\\
			112 &	 Luiz Bonfa- Manha De Carnaval\\
			113 &	 Marcos Valle- Preciso Aprender A Viver So\\
			114 &	 Marisa Monte- Ainda Lembro\\
			115 &	 Marisa Monte- Amor I Love You\\
			116 &	 Marisa Monte- Ando Meio Desligado\\
			117 &	 Tom Jobim- Aguas De Marco\\
			118 &	 Tom Jobim- Amor E Paz\\
			119 &	 Tom Jobim- Brigas Nunca Mais\\
			120 &	 Tom Jobim- Desafinado\\
			121 &	 Tom Jobim- Fotografia\\
			122 &	 Tom Jobim- Garota De Ipanema\\
			123 &	 Tom Jobim- Meditacao\\
			124 &	 Tom Jobim- Samba Do Aviao\\
			125 &	 Tom Jobim- Se Todos Fossem Iguais A Voce\\
			126 &	 Tom Jobim- So Tinha De Ser Com Voce\\
			127 &	 Tom Jobim- Vivo Sonhando\\
			128 &	 Tom Jobim- Voce Abusou\\
			129 &	 Tom Jobim- Wave\\
			130 &	 Toquinho- Agua Negra Da Lagoa\\
			131 &	 Toquinho- Ao Que Vai\\
			132 &	 Toquinho- Este Seu Olhar\\
			133 &	 Vinicius De Moraes- Apelo\\
			134 &	 Vinicius De Moraes- Carta Ao Tom\\
			135 &	 Vinicius De Moraes- Minha Namorada\\
			136 &	 Vinicius De Moraes- O Morro Nao Tem Vez\\
			137 &	 Vinicius De Moraes- Onde Anda Voce\\
			138 &	 Vinicius De Moraes- Pela Luz Dos Olhos Teus\\
			139 &	Vinicius De Moraes- Samba Em Preludio\\
			140 &	Vinicius De Moraes-Tereza Da Praia\\
        \hline\hline
\end{tabular}
\end{indented}
\end{minipage}
\label{tab:BluesAndBossa}
\end{table}

\begin{table}
\caption{Reggae and Rock art works}
\begin{minipage}[b]{.30\textwidth}
\begin{indented}
       \item[]
       \tiny
       \color[rgb]{0,0.7,0.7}
       \arrayrulecolor{black}
        \begin{tabular}{l c c}
        \hline
          & \textcolor{black}{\textbf{Reggae art works}}  \\
        \hline\hline
        141 &    Ace Of Bass- All That She Wants\\
        142 &    Ace Of Bass- Dont Turn Around\\
				143 &		 Ace Of Bass- Happy Nation\\
				144 &		 Armandinho- Pela Cor Do Teu Olho\\
				145 &		 Armandinho- Sentimento\\
				146 &		 Big Mountain- Baby I Love Your Way\\
				147 &		 Bit Meclean- Be Happy\\
				148 &		 Bob Marley- Africa Unite\\
				149 &		 Bob Marley- Buffalo Soldier\\
				150 &		 Bob Marley- Exodus\\
				151 &		 Bob Marley- Get Up Stand Up\\
				152 &		 Bob Marley- I Shot The Sheriff\\
				153 &		 Bob Marley- Iron Lion Zion\\
				154 &		 Bob Marley- Is This Love\\
				155 &		 Bob Marley- Jammin\\
				156 &		 Bob Marley- No Woman No Cry\\
				157 &		 Bob Marley- Punky Reggae Party\\
				158 &		 Bob Marley- Root Rock\\
				159 &		 Bob Marley- Satisfy\\
				160 &		 Bob Marley- Stir It Up\\
				161 &		 Bob Marley- Three Little Birds\\
				162 &		 Bob Marley- Waiting In The Van\\
				163 &		 Bob Marley- War\\
				164 &		 Bob Marley- Wear My\\
				165 &		 Bob Marley- Zimbabwe\\
				166 &		 Cidade Negra- A Cor Do Sol\\
				167 &		 Cidade Negra- A Flexa E O Vulcao\\
				168 &		 Cidade Negra- A Sombra Da Maldade\\
				169 &		 Cidade Negra- Aonde Voce Mora\\
				170 &		 Cidade Negra- Eu Fui Eu Fui\\
				171 &		 Cidade Negra- Eu Tambem Quero Beijar\\
				172 &		 Cidade Negra- Firmamento\\
				173 &		 Cidade Negra- Girassol\\
				174 &		 Cidade Negra- Ja Foi\\
				175 &		 Cidade Negra- Mucama\\
				176 &		 Cidade Negra- O Ere\\
				177 &		 Cidade Negra- Pensamento\\
				178 &		 Dazaranhas- Confesso\\
				179 &		 Dont Worry\\
				180 &		 Flor Do Reggae\\
				181 &		 Inner Circle- Bad Boys\\
				182 &		 Inner Circle- Sweat\\
				183 &		 Jimmy Clif- I Can See Clearly Now\\
				184 &		 Jimmy Clif- Many Rivers To Cross\\
				185 &		 Jimmy Clif- Reggae Night\\
				186 &		 Keep On Moving\\
				187 &		 Manu Chao- Me Gustas Tu\\
				188 &		 Mascavo- Anjo Do Ceu\\
				189 &		 Mascavo- Asas\\
				190 &		 Natiruts- Liberdade Pra Dentro Da Cabeca\\
				191 &		 Natiruts- Presente De Um Beija Flor\\
				192 &		 Natiruts- Reggae Power\\
				193 &		 Nazarite Skank\\
				194 &		 Peter Tosh- Johnny B Goode\\
				195 &		 Shaggy- Angel\\
				196 &		 Shaggy- Bombastic\\
				197 &		 Shaggy- It Wasnt Me\\
				198 &		 Shaggy- Strenght Of A Woman\\
				199 &		 Sublime- Badfish\\
				200 &		 Sublime- D Js\\
				201 &		 Sublime- Santeria\\
				202 &		 Sublime- Wrong Way\\
				203 &		 Third World- Now That We Found Love\\
				204 &		 Tribo De Jah- Babilonia Em Chamas\\
				205 &		 Tribo De Jah- Regueiros Guerreiros\\
				206 &		 Tribo De Jah- Um So Amor\\
				207 &		 U B40- Bring Me Your Cup\\
				208 &		 U B40- Home Girl\\
				209 &		 U B40- To Love Somebody\\
				210 &		 Ub40- Red Red Wine		\\
        \hline\hline
\end{tabular}
\end{indented}
\end{minipage}\qquad
\begin{minipage}[b]{.30\textwidth}
       \begin{indented}
       \item[]
       \tiny
       \color[rgb]{0.5, 0, 1}     
       \begin{tabular}{l c c}
        \hline
          & \textcolor{black}{\textbf{Rock art works}}  \\
        \hline\hline
        211 &    Aerosmith- Kings And Queens\\
				212 &	     Aha- Stay On These Roads\\
				213 &	 Aha- Take On Me\\
				214 &	 Aha- Theres Never A Forever Thing\\
				215 &	 Beatles- Cant Buy Me Love\\
				216 &	 Beatles- From Me To You\\
				217 &	 Beatles- I Want To Hold Your Hand\\
				218 &	 Beatles- She Loves You\\
				219 &	 Billy Idol- Dancing With Myself\\
				220 &	 Cat Steves- Another Saturday Night\\
				221 &	 Creed- My Own Prison\\
				222 &	 Deep Purple- Deamon Eye\\
				223 &	 Deep Purple- Hallelujah\\
				224 &	 Deep Purple- Hush\\
				225 &	 Dire Straits- Sultan Of Swing\\
				226 &	 Dire Straits- Walk Of Life\\
				227 &	 Duran Duran- A View To A Kill\\
				228 &	 Eric Clapton- Cocaine\\
				229 &	 Europe- Carrie\\
				230 &	 Fleetwood Mac- Dont Stop\\
				231 &	 Fleetwood Mac- Dreams\\
				232 &	 Fleetwood Mac- Gold Dust Woman\\
				233 &	 Foo Fighters- Big Me\\
				234 &	 Foo Fighters- Break Out\\
				235 &	 Foo Fighters- Walking After You\\
				236 &	 Men At Work- Down Under\\
				237 &	 Men At Work- Who Can It Be Now\\
				238 &	 Metallica- Battery\\
				239 &	 Metallica- Fuel\\
				240 &	 Metallica- Hero Of The Day\\
				241 &	 Metallica- Master Of Puppets\\
				242 &	 Metallica- My Friendof Misery\\
				243 &	 Metallica- No Leaf Clover\\
				244 &	 Metallica- One\\
				245 &	 Metallica- Sad But True\\
				246 &	 Pearl Jam- Alive\\
				247 &	 Pearl Jam- Black\\
				248 &	 Pearl Jam- Jeremy\\
				249 &	 Pet Shop Boys- Go West\\
				250 &	 Pet Shop Boys- One In A Million\\
				251 &	 Pink Floyd- Astronomy Domine\\
				252 &	 Pink Floyd- Have A Cigar\\
				253 &	 Pink Floyd- Hey You\\
				254 &	 Queen- Another One Bites The Dust\\
				255 &	 Queen- Dont Stop Me Now\\
				256 &	 Queen- I Want It All\\
				257 &	 Queen- Play The Game\\
				258 &	 Queen- Radio Gaga\\
				259 &	 Queen- Under Pressure\\
				260 &	 Red Hot Chilli Pepers- Higher Ground\\
				261 &	 Red Hot Chilli Pepers- Otherside\\
				262 &	 Red Hot Chilli Pepers- Under The Bridge\\
				263 &	 Rolling Stones- Angie\\
				264 &	 Rolling Stones- As Tears Go By\\
				265 &	 Rolling Stones- Satisfaction\\
				266 &	 Rolling Stones- Street Of Love\\
				267 &	 Steppen Wolf- Magic Carpet Ride\\
				268 &	 Steve Winwood- Valerie\\
				269 &	 Steve Winwood- While You See A Chance\\
				270 &	 Tears For Fears- Shout\\
				271 &	 The Doors- Hello I Love You\\
				272 &	 The Doors- Light My Fire\\
				273 &	 The Doors- Love Her Madly\\
				274 &	 U2- Elevation\\
				275 &	 U2- Ever Lasting Love\\
				276 &	 U2- When Love Comes To Town\\
				277 &	 Van Halen- Dance The Night Away\\
				278 &	 Van Halen- Dancing With The Devil\\
				279 &	 Van Halen- Jump\\
				280 &	 Van Halen- Panama\\
        \hline\hline
\end{tabular}
\end{indented}
\end{minipage}
  \label{tab:ReggaeAndRock}
\end{table}

\begin{figure}
\subfigure[]
{
   \centering
   \includegraphics[width=12cm]{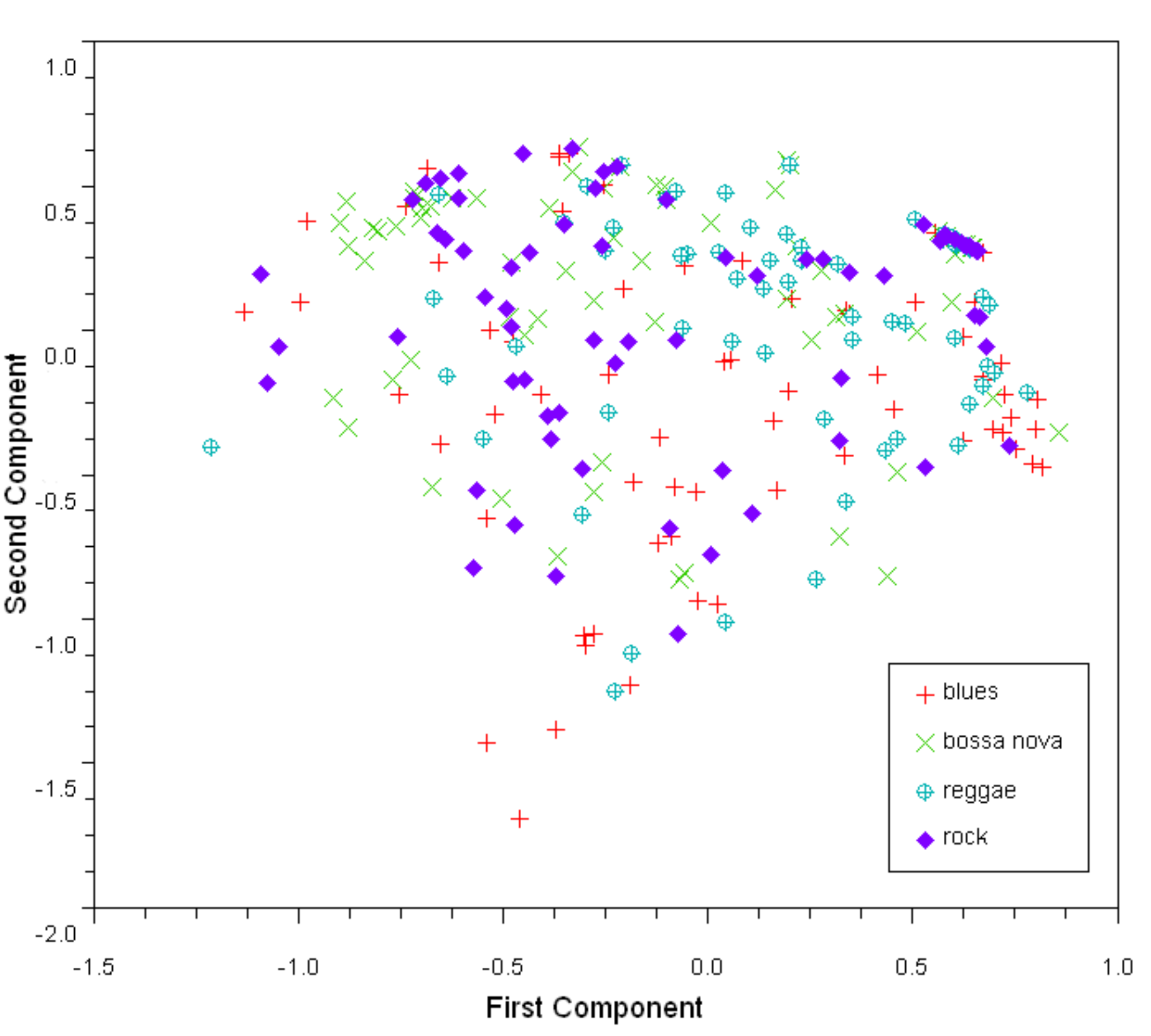}
   \label{fig:pca12}
}
\hspace{0.3cm}
\subfigure[]
{
   \centering
   \includegraphics[width=12cm]{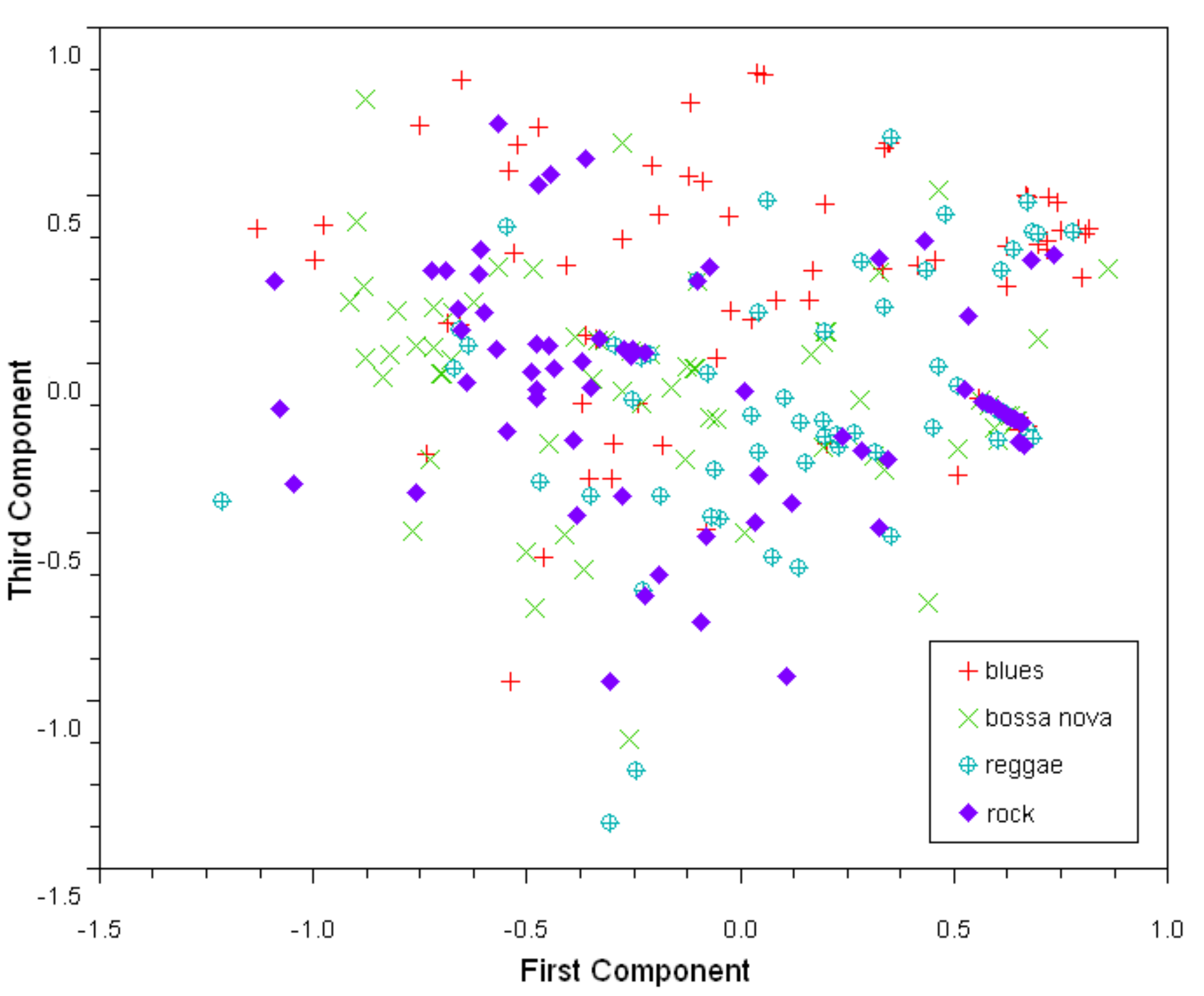}
   \label{fig:pca13}
}
\caption{The new first three feaures obtained by PCA. (a) First and second axes. (b) First and third axes.}
\label{fig:pca}
\end{figure}

Once we want to reduce data dimensionality, it is necessary to set the
suitable number of principal components that will represent the new
features. Not surprisingly, in a high dimensional space the classes
can be easily separated. On the other hand, high dimensionality
increases complexity, making the analysis of both extracted features
and classification results a difficult task. One approach to get the
ideal number of principal components is to verify how much of the
data variance is preserved. In order to do so, the $l$ first
eigenvalues ($l$ is the quantity of principal components to be
verified) are summed up and the result is divided by the the sum of
all the eigenvalues. If the result of this calculation is a value
equal or greater than 0.75, it is said that these number of components
(or new features) preserves at least 75\% of the data variance, which
is often enough for classification purposes. When PCA was applied to
the normalized rythm features, as illustrated in Figure
\ref{fig:pca}, it was observed that 20 principal components
preserved 76\% of the variance of the data. That is, it is possible to
reduce the data dimensionality from 364-D to 20-D without a
significant loss of information. Nevertheless, as will be shown in the
following, depending on the classifier and how the classification task
was performed, different number of components were required in each
situation in order to achieve suitable results. Despite the fact of
preserving only 32\% of the variance, Figure \ref{fig:pca} shows the
first three principal components, that is, the first three new
features obtained with PCA. Figure
\ref{fig:pca12} shows the first and second features and Figure
\ref{fig:pca13} shows the first and third features. It is noted that
the classes are completely overlapped, making the problem of automatic
classification a nontrivial task.

\clearpage

In all the following supervised classification tasks, re-substitution
means that all objects from each class were used as the training set
(in order to estimate the parameters) and all objects were used as the
testing set. Hold-out 70\%-30\% means that 70\% of the objects from
each class were used as the training set and 30\% (different ones) for
testing. Finally, in hold-out 50\%-50\%, the objects were separated
into two groups: 50\% for training and 50\% for testing.

The kappa variance is strongly  related to its  accuracy, that is, how
reliable are its  value.   The higher is its  variance, the lower is its
accuracy. Once the kappa coefficient is  a statistics, in general, the
use  of large datasets  improves its accuracy  by  making its variance
smaller.  This concept can   be observed in   the results. The smaller
variance occurred   in the  re-substitution  situation,  in which  all
samples     constitute the  testing    set.     This indicates    that
re-substitution   provided     the  best  kappa    accurary    in  the
experiments. On the other hand, hold-out 70\%-30\% provided the higher
kappa variance, once only 30\% of  the samples establishes the testing
set.

The results obtained by the Quadratic Bayesian Classifier using PCA
are shown in Table \ref{tab:PCAQBayes} in terms of kappa, its
variance, the accurary of the classification and the overall
performance according to the value of kappa.

\begin{table}
    \centering    
    \caption{PCA kappa and accuracy for the Quadratic Bayesian Classifier}
    \footnotesize
        \begin{tabular}{l c c c c c}
        \hline
        & \textbf{Kappa} & \textbf{Variance} & \textbf{Accuracy} & \textbf{Performance}  \\
        \hline\hline
        Re-Substitution & 0.66 & 0.0012 & 74.65\% & Substantial\\ 
        Hold-Out (70\%-30\%) & 0.23 & 0.0057 & 42.5\% & Fair\\ 
        Hold-Out (50\%-50\%) & 0.22 & 0.0033 & 41.43\% & Fair \\ 
        \hline\hline
        \end{tabular}
        \label{tab:PCAQBayes}
\end{table}

Table \ref{tab:PCAQBayes} also indicates that the performance was not
satisfactory for the Hold-Out (70\%-30\%) and Hold-Out (50\%-50\%). As
the PCA is a not supervised approach, the parameter estimation
performance (of covariance matrices for instance) is strongly degraded
because of the small sample size problem.

The confusion matrix for the re-substitution classification task in
Table \ref{tab:PCAQBayes} is illustrated in Table
\ref{tab:MatConfPCA}. All reggae samples were classified correctly. In
addition, many samples from the other classes were classified as
reggae. 

\begin{table}
    \centering
    \caption{Confusion Matrix for the Quadratic Bayesian Classifier using PCA and Re-Substitution}
    \footnotesize
        \begin{tabular}{l c c c c c}
        \hline
        &  Blues & \textit{Bossa nova} & Reggae & Rock  \\
        \hline\hline
        Blues      & 48 & 0  & 22 &  0 \\ 
        \textit{Bossa nova} & 2 & 47  & 21 &  0 \\ 
        Reggae     & 0 &  0  & 70 &  0 \\ 
        Rock       & 1 &  1  & 24 & 44 \\ 
        \hline\hline
        \end{tabular}
        \label{tab:MatConfPCA}
\end{table}

Table \ref{tab:WrongsPCA} presents the misclassified art works of the
confusion matrix in Table \ref{tab:MatConfPCA}.

\begin{table}
\caption{Misclassified art works using PCA with Quadratic Bayesian Classifier and Re-Substitution}
\scriptsize
        \begin{tabular}{l c c}
        \hline
        Blues as Reggae & 3 6 10 12 16 17 18 23 25 28 34 36 37 38 40 51 52 54 56 58 63 69\\
        \textit{Bossa nova} as Blues & 123 126\\
        \textit{Bossa nova} as Reggae & 74 82 83 86 89 96 97 98 100 103 104 105 106 107 109 112 114 115 117 118 125\\
        Rock as Blues & 266\\
        Rock as \textit{Bossa nova} & 212\\
        Rock as Reggae & 217 219 226 228 231 233 236 237 239 245 247 249 254 255 257 261 262 264 267 273 274 275 276 278\\
        \hline\hline
\end{tabular}
 \label{tab:WrongsPCA}
\end{table}

With the purpose of comparing two different classifiers, the obtained
results by the Linear Bayesian Classifier using PCA are shown in Table
\ref{tab:PCALBayes}, again, in terms of kappa, its variance, the
accurary of the classification and the overal performance according to
its value. The performance of the Hold-Out (70\%-30\%) and Hold-Out
(50\%-50\%) classification task increased slightly, mainly due to the
fact that here a unique covariance matrix is estimated using all the
samples in the dataset.

\begin{table}
    \centering
    \caption{PCA kappa and accuracy for the Linear Bayesian Classifier}
    \footnotesize
        \begin{tabular}{l c c c c c}
        \hline
        & \textbf{Kappa} & \textbf{Variance} & \textbf{Accuracy} & \textbf{Performance}  \\
        \hline\hline
        Re-Substitution & 0.63 & 0.0013 & 72.15\% & Substantial\\ 
        Hold-Out (70\%-30\%) & 0.35 & 0.0055 & 51.25\% & Fair\\ 
        Hold-Out (50\%-50\%) & 0.31 & 0.0032 & 48.58\% & Fair \\ 
        \hline\hline
        \end{tabular}
        \label{tab:PCALBayes}
\end{table}

\begin{figure}
\centering
\subfigure[]
{
    \label{fig:KappaReSubQuad:a}
    \includegraphics[width=7.2cm]{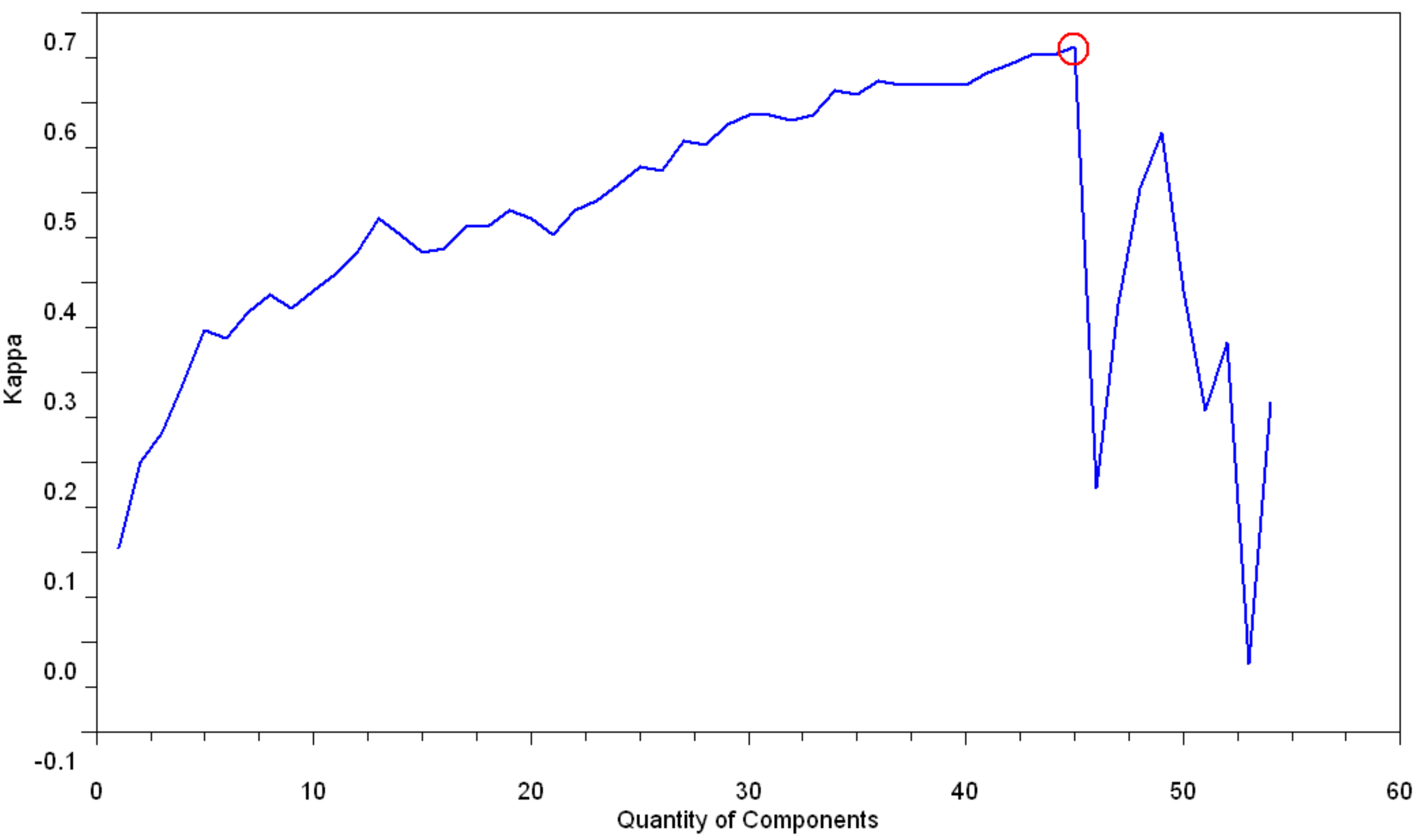}
}
\hspace{0.3cm}
\subfigure[]
{
    \label{fig:KappaReSubLinear:b}
    \includegraphics[width=7.2cm]{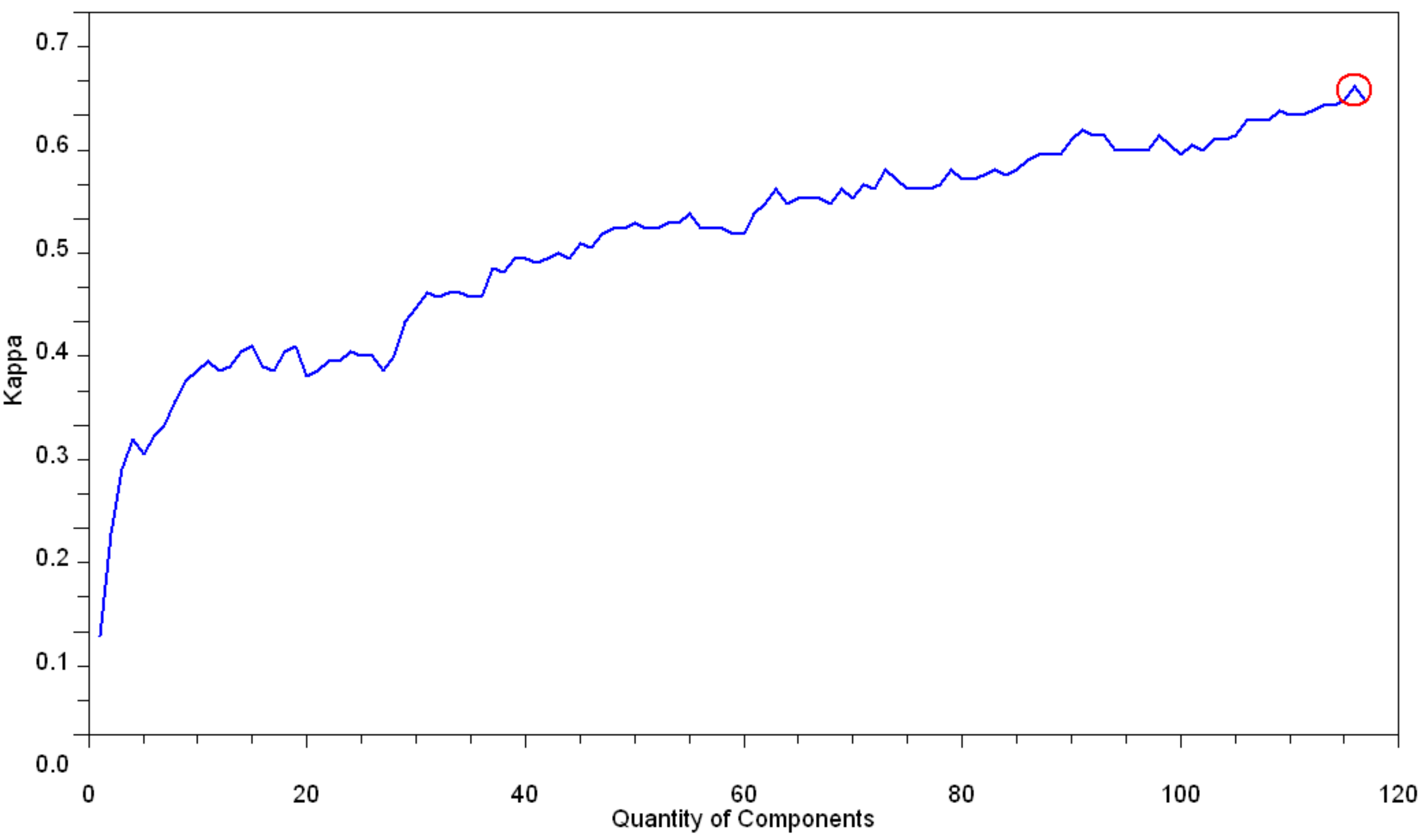}
}
\hspace{0.3cm}
\subfigure[]
{
    \label{fig:Kappa7030Quad:c}
    \includegraphics[width=7.2cm]{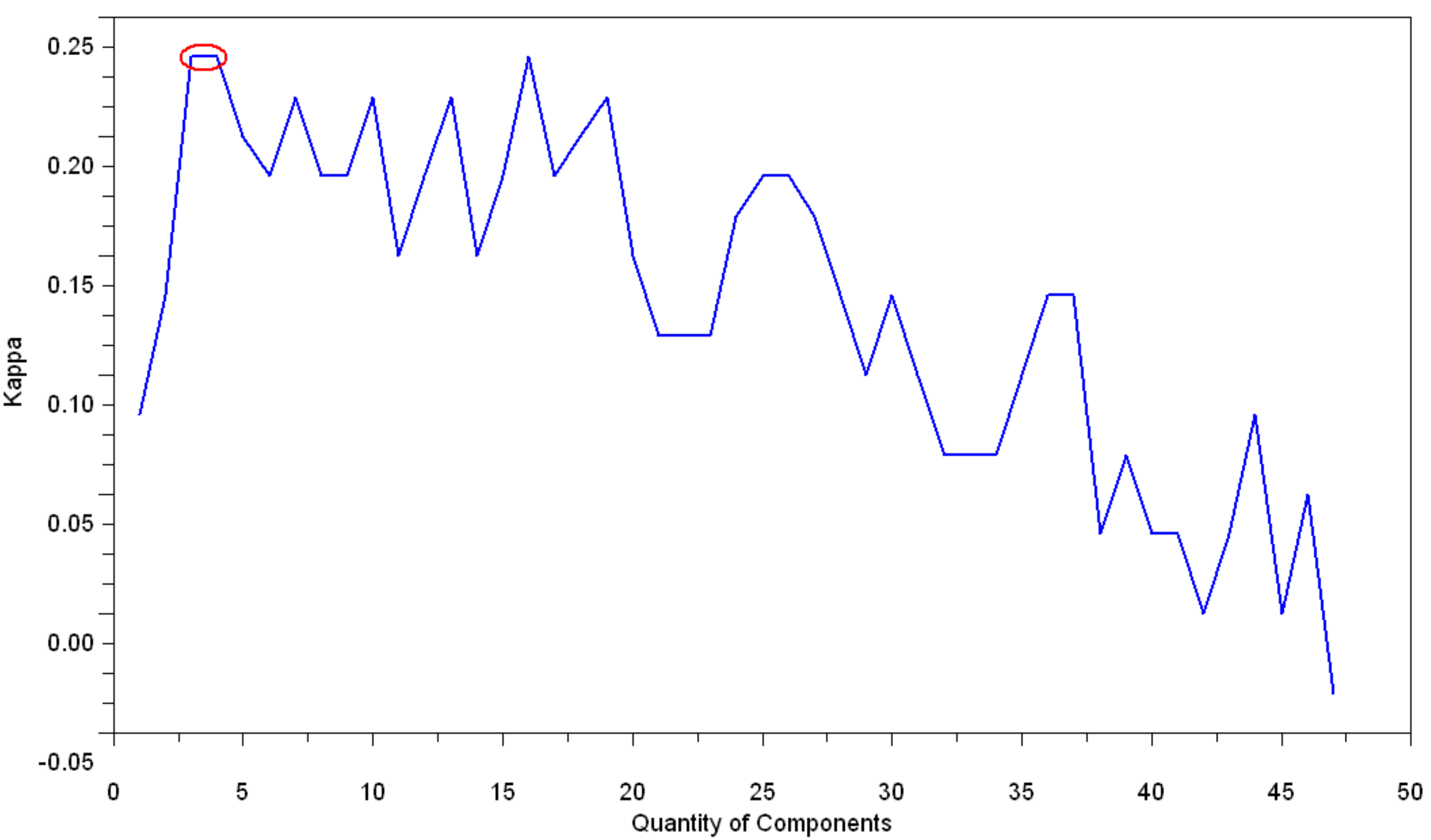}
}
\hspace{0.3cm}
\subfigure[]
{
    \label{fig:Kappa7030Linear:d}
    \includegraphics[width=7.2cm]{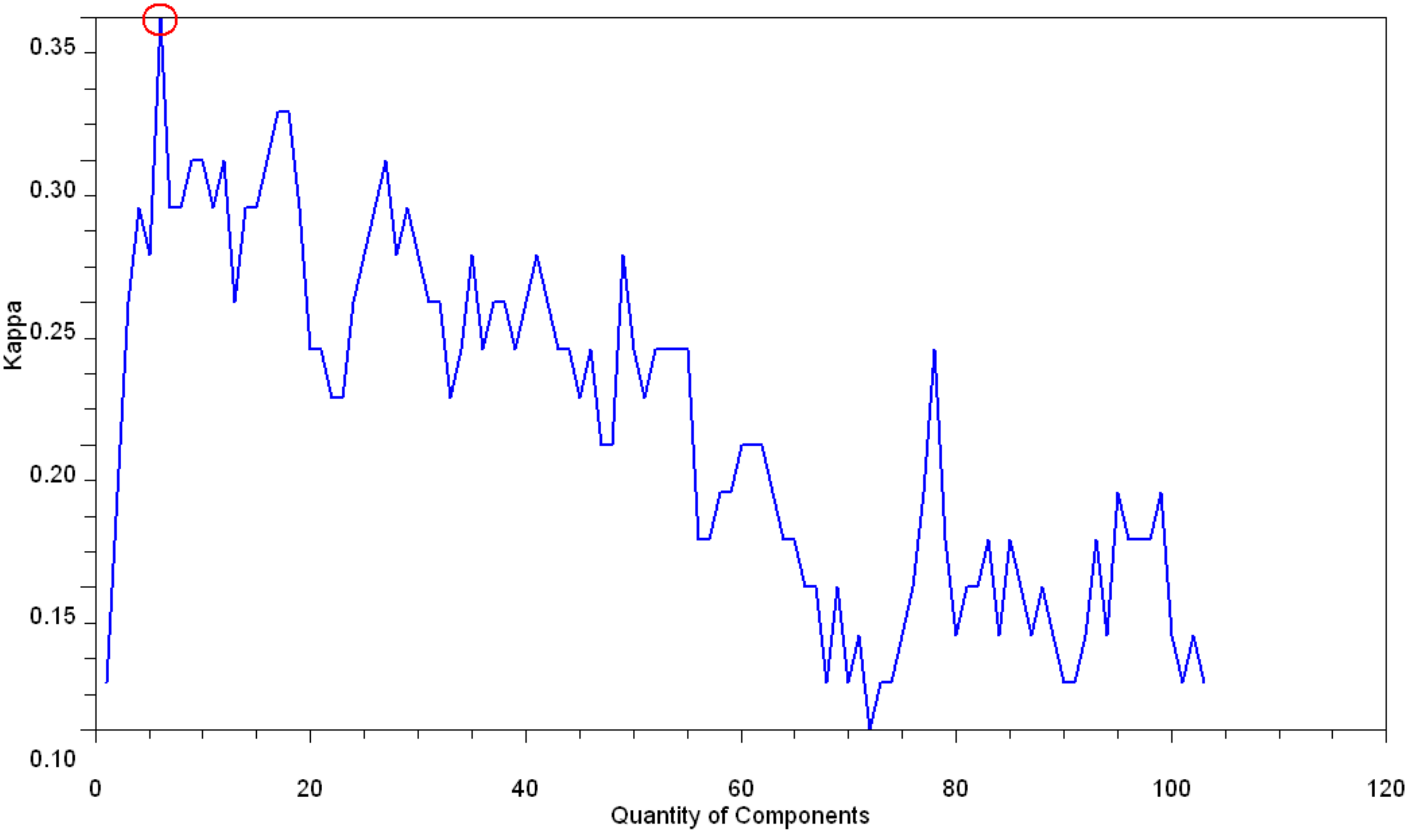}
}
\hspace{0.3cm}
\subfigure[]
{
    \label{fig:Kappa5050Quad:e}
    \includegraphics[width=7.2cm]{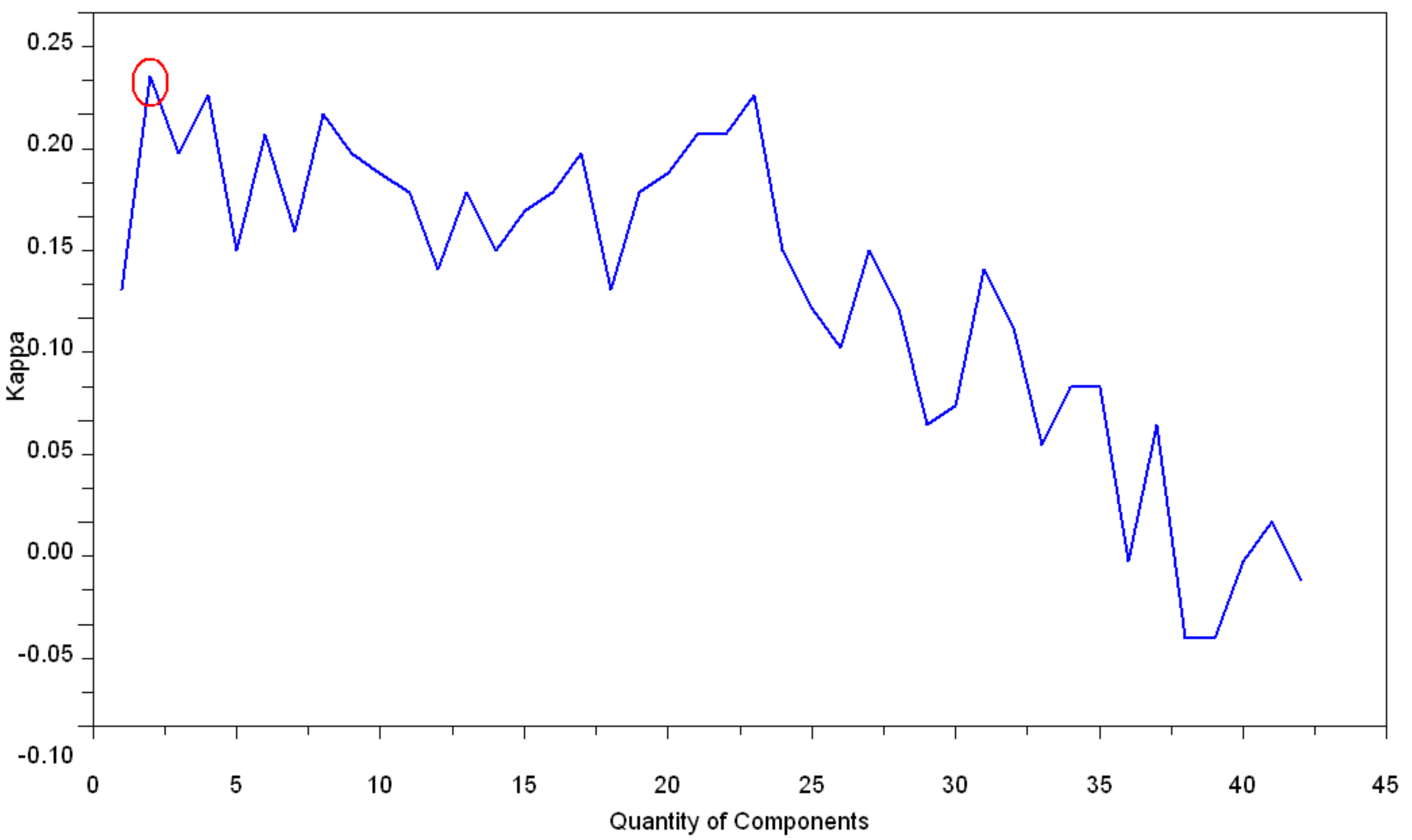}
}
\hspace{0.3cm}
\subfigure[]
{
    \label{fig:Kappa5050Linear:f}
    \includegraphics[width=7.2cm]{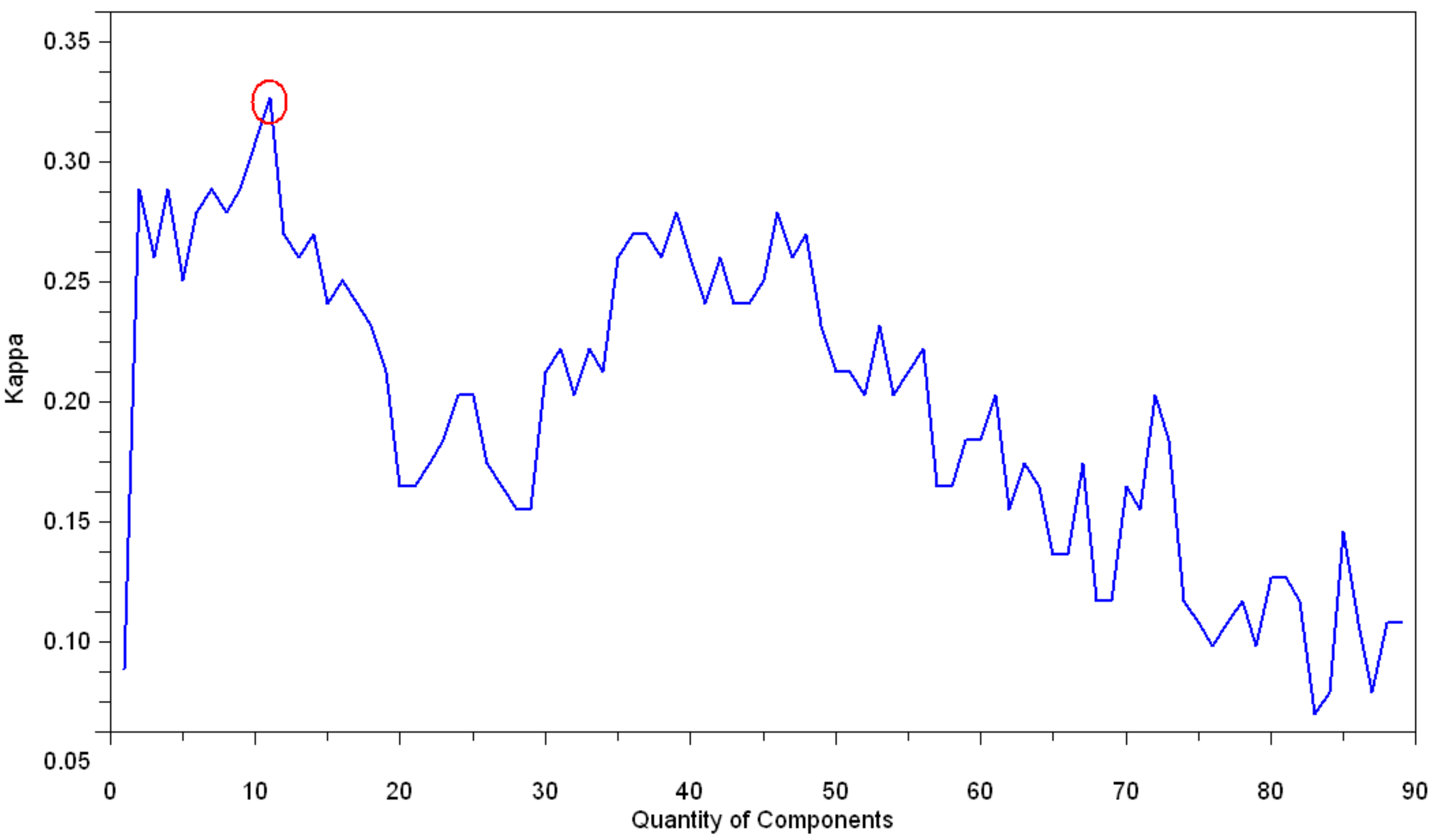}
}
  \caption{Values of Kappa varying the number of principal components
  (a)Quadratic Bayesian Classifier: Re-Substitution (b)Linear Bayesian
  Classifier: Re-Substitution (c)Quadratic Bayesian Classifier:
  Hold-Out (70\%-30\%) (d)Linear Bayesian Classifier: Hold-Out
  (70\%-30\%) (e)Quadratic Bayesian Classifier: Hold-Out (50\%-50\%)
  (f)Linear Bayesian Classifier: Hold-Out (50\%-50\%).}
\label{fig:Kappa}

\end{figure}

Figure \ref{fig:Kappa} depicts the value of kappa depending on the
quantity of principal components used in the Quadratic and Linear
Bayesian Classifier. The last value of each graphic makes it clear
that from this value onwards the classification can not be done due to
singularity problems involving the inversion of the covariance
matrices (curse of dimensionality). It can be observed that this
singularity threshold is different in each situation. However, for the
quadratic classifier this value is in the range of about 40-55
components; while, for the linear classifier, is in the range of about
86-115 components. The smaller quantity for the quadratic classifier
can be explained by the fact that there are four covariance matrices,
each one estimated from the samples for one respective class. As there
are 70 samples in each class, singularity problems will occur in a
smaller dimensional space when compared to the linear classifier, that
uses all the 280 samples to estimate one unique covariance
matrix. Therefore, the ideal number of principal components allowing
the highest value of kappa should be those circled in red in Figure
\ref{fig:Kappa}.

Keeping in mind that the problem of automatic genre classification is
a nontrivial task, that in this study only one aspect of the rhythm is
been analysed (the occurrence of the ryhthm notations), and that PCA
is an unsupervised approach for feature extraction, the correct
classifications presented in Tables \ref{tab:PCAQBayes} and
\ref{tab:PCALBayes} for the re-substitution situation corroborate
strongly the viability of the proposed methodology. In spite of the
complexity of comparing differents proposed approaches to automatic
genre classification discussed in the Introduction, these accuracy
values are very close or even superior when compared to previous works
\cite{MOSTAFA2009, WANG2008, SONG2008, HOLZAPFEL2007, SCARINGELLA2006}.

\begin{figure}
\subfigure[]
{
   \centering
   \includegraphics[width=12cm]{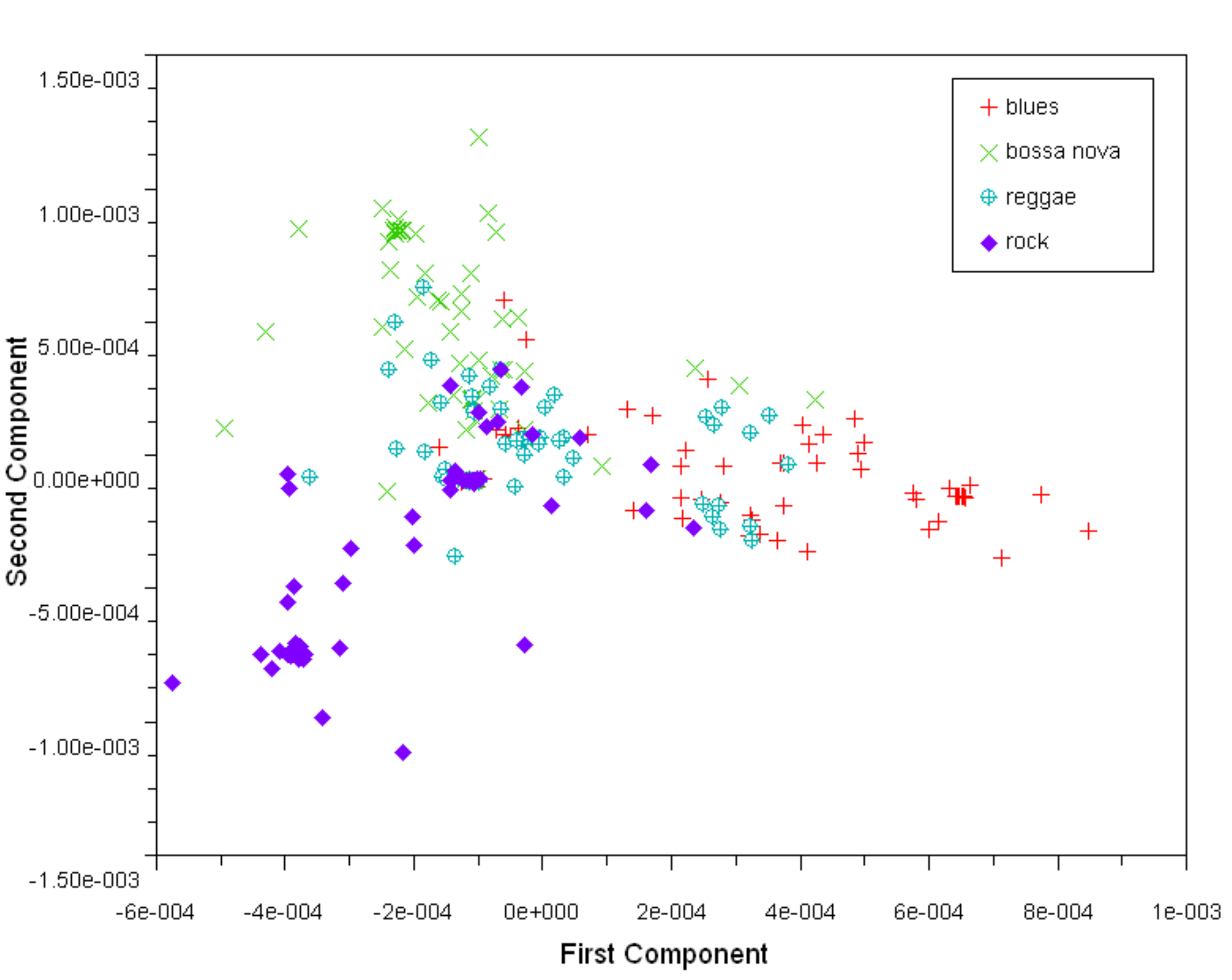}
   \label{fig:lda12}
}
\hspace{0.3cm}
\subfigure[]
{
   \centering
   \includegraphics[width=12cm]{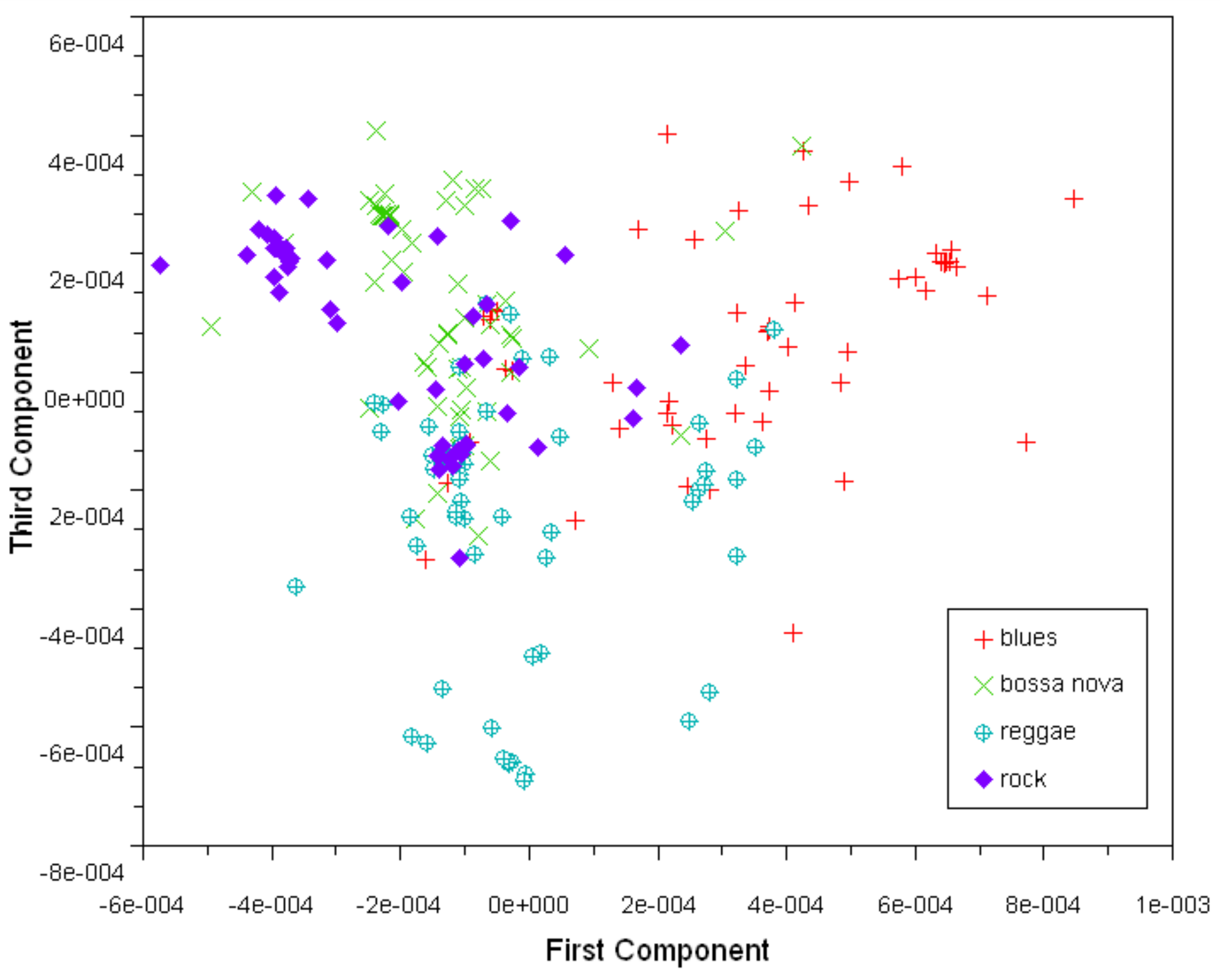}
   \label{fig:lda13}
}
\caption{The new first three feaures obtained by LDA. (a) First and second features. (b) First and third features.}
\label{fig:lda}
\end{figure}

Similarly, Figure \ref{fig:lda} shows the three components, namely the
three new features obtained with LDA. As mentioned before, LDA
approach has a restriction of obtaining only $C-1$ nonzero
eigenvalues, where $C$ is the number of class. Therefore, only three
components are computed. Once it is a supervised approach and the main
goal is to maximize class separability, the four classes in Figure
\ref{fig:lda12} and Figure \ref{fig:lda13} are clearer than in PCA,
although still involving substantial overlaps. This result
corroborates that automatic classification of musical genres is not a
trivial task.

Table \ref{tab:LDAQBayes} presents the results obtained by the
Quadratic Bayesian Classifer using LDA. Differently from PCA, the use
of hold-out (70\%-30\%) and hold-out (50\%-50\%) provided good
results, what is notable and reflects the supervised characteristic of
LDA, which makes use of all discriminant information available in the
feature matrix.

\begin{table}
    \centering
    \caption{LDA kappa and accuracy for the Quadratic Bayesian Classifier}
    \footnotesize
        \begin{tabular}{l c c c c c}
        \hline
        & \textbf{Kappa} & \textbf{Variance} & \textbf{Accuracy} & \textbf{Performance}  \\
        \hline\hline
        Re-Substitution & 0.66 & 0.0012 & 74.29\% & Substantial\\ 
        Hold-Out (70\%-30\%) & 0.62 & 0.0045 & 71.25\% & Substantial\\ 
        Hold-Out (50\%-50\%) & 0.64 & 0.0025 & 72.86\% & Substantial \\ 
        \hline\hline
        \end{tabular}
        \label{tab:LDAQBayes}
\end{table}

Despite the value of kappa and its variance being the same using LDA
with re-substitution and PCA with re-substitution, the two confusion
matrix are strongly distinct each other. In the first case,
demostrated in Table \ref{tab:MatConfLDA}, the misclassified art works
are well distributed among the four classes, while with PCA (Table
\ref{tab:MatConfPCA}) they are concentrated in one class, represented
by the reggae genre. The results obtained with LDA technique are
particularly promising because they reflect the nature of the
data. Although widely used, terms such as rock, reggae or pop often
remain loosely defined \cite{SCARINGELLA2006}. Yet, it is worthwhile
to remember that the intensity of the beat, which is a very important
aspect of the rhythm has not been considered in this work. This means
that analysing rhythm only through notations, as currently proposed,
could poise difficulties even for human experts. These misclassified
art works have similar properties described in terms of rhythm
notations and, as a result, they generate similar weight
matrices. Therefore, the proposed methodology, although requiring some
complementations, seems to be a significant contribution towards the
development of viable alternative approach to automatic genre
classification.

\begin{table}
    \centering
    \caption{Confusion Matrix for the Quadratic Bayesian Classifier using LDA and Re-Substitution}
    \footnotesize
        \begin{tabular}{l c c c c c}
        \hline
        &  Blues & \textit{Bossa nova} & Reggae & Rock  \\
        \hline\hline
        Blues      & 58 & 6  & 4 &  2 \\ 
        \textit{Bossa nova} & 3 & 52  & 7 &  8 \\ 
        Reggae     & 8 &  8  & 44 &  10 \\ 
        Rock       & 4 &  7  & 5 & 54 \\ 
        \hline\hline
        \end{tabular}
        \label{tab:MatConfLDA}
\end{table}

The misclassified art works of the confusion matrix in Table
\ref{tab:MatConfLDA} are identified in Table \ref{tab:WrongsLDA}.

\begin{table}
\caption{Misclassified art works using LDA with Quadratic Bayesian Classifier and Re-Substitution}
\centering
\scriptsize
        \begin{tabular}{l c c}
        \hline
        Blues as \textit{Bossa nova} & 2 30 52 54 58 63\\
        Blues as Reggae & 5 17 28 69\\
        Blues as Rock & 6 51\\
        \textit{Bossa nova} as Blues & 79 80 116\\
        \textit{Bossa nova} as Reggae & 90 96 97 105 112 122 131\\
        Bossa Bova as Rock & 76 78 83 98 100 103 107 109 \\
        Reggae as Blues & 144 152 166 170 188 190 193 203 \\
        Reggae as \textit{Bossa nova} & 158 159 162 164 169 181 197 208 \\
        Reggae as Rock & 146 156 157 160 168 175 180 189 198 210\\
        Rock as Blues & 238 239 245 278\\
        Rock as \textit{Bossa nova} & 233 237 254 262 264 266 276 \\
        Rock as Reggae & 219 228 255 261 263\\
        \hline\hline
\end{tabular}
 \label{tab:WrongsLDA}
\end{table}

The results for the Linear Bayesian Classifier using LDA are shown in
Table \ref{tab:LDALBayes}. In fact, they are closely similar to those
obtained by the Quadratic Bayesian Classifier (Table
\ref{tab:LDAQBayes}).

\begin{table}
    \centering
    \caption{LDA kappa and accuracy for the Linear Bayesian Classifier}
    \footnotesize
        \begin{tabular}{l c c c c c}
        \hline
        & \textbf{Kappa} & \textbf{Variance} & \textbf{Accuracy} & \textbf{Performance}  \\
        \hline\hline
        Re-Substitution & 0.65 & 0.0012 & 73.58\% & Substantial\\ 
        Hold-Out (70\%-30\%) & 0.63 & 0.0043 & 72.5\% & Substantial\\ 
        Hold-Out (50\%-50\%) & 0.64 & 0.0025 & 72.86\% & Substantial \\ 
        \hline\hline
        \end{tabular}
        \label{tab:LDALBayes}
\end{table}

As mentioned in section \ref{subsec:LDA}, linear discriminant analysis
also allows us to quantify the intra and interclass dispersion of the
feature matrix through functionals such as the trace and determinant
computed from the scatter matrices \cite{FUKUNAGA1990}. The overall
intraclass scatter matrix, hence $S_{intra}$; the intraclass scatter
matrix for each class, hence $S_{intraBlues}$, $S_{intraBossaNova}$,
$S_{intraReggae}$ and $S_{intraRock}$; the interclass scatter matrix,
hence, $S_{inter}$; and the overall separability index, hence
$\left(S_{intra}^{-1}*S_{inter}\right)$, were computed. Their
respective traces are:

\begin{center}
$trace(S_{intra})$ = \qquad 499.526 \\
$trace(S_{intraBlues})$ = \qquad  138.615 \\
$trace(S_{intraBossaNova})$ = \qquad   119.302 \\
$trace(S_{intraReggae})$ = \qquad   98.327 \\
$trace(S_{intraRock})$ = \qquad   143.280 \\ 
$trace(S_{inter})$ = \qquad 21.598 \\
$trace\left(S_{intra}^{-1}*S_{inter}\right)$ = \qquad 3.779 \\
\end{center}

Two important observations are worth mentioning. First, these traces
emphasise the difficulty of this classification problem: the traces of
the intraclass scatter matrices are too high, and the trace of the
interclass scatter matrix together with the overal separability index,
too small. This confirms that the four classes are overlapping
completely. Second, the smaller intraclass trace is related to the
reggae genre (it is the most compact class). This may justify why in
the experiments art works belonging to reggae were more frequently
90\%-100\% correctly classified.

\begin{figure}
\subfigure[]
{
   \centering
   \includegraphics[width=7cm]{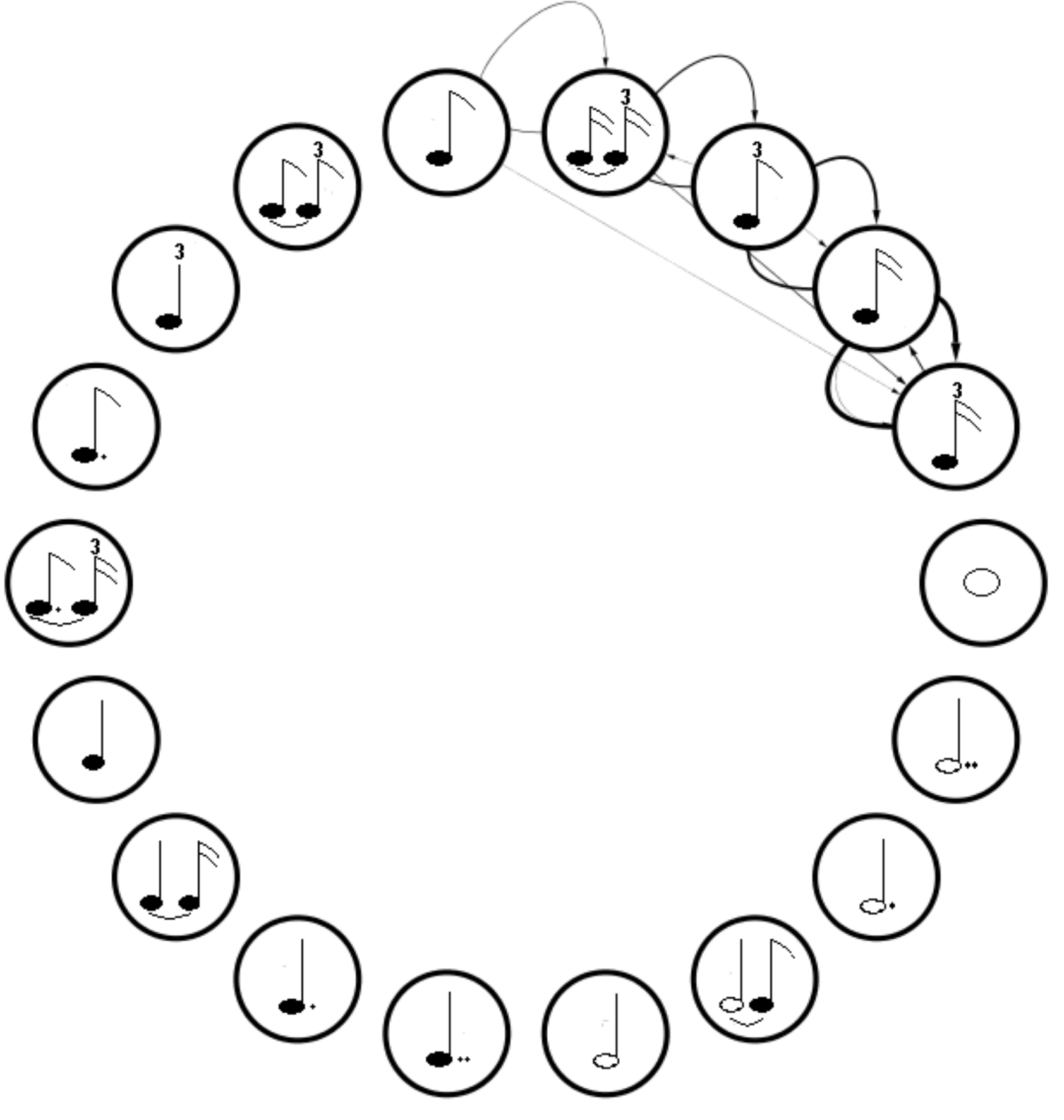}
   \label{fig:PCAAutoVetor1}
}
\hspace{0.3cm}
\subfigure[]
{
   \centering
   \includegraphics[width=7cm]{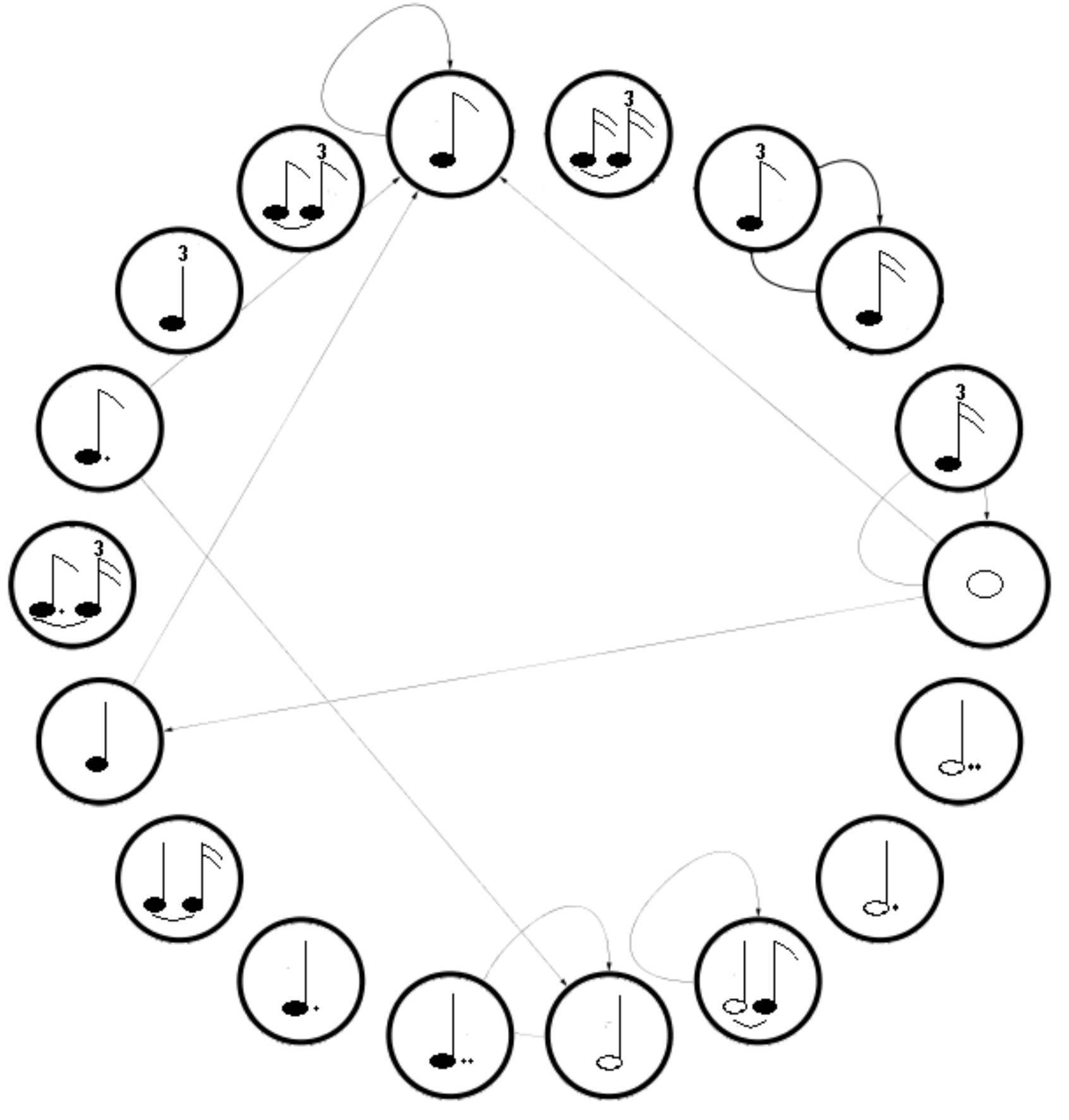}
   \label{fig:PCAAutoVetor2}
} 
\hspace{0.3cm}
\subfigure[]
{
   \centering
   \includegraphics[width=7cm]{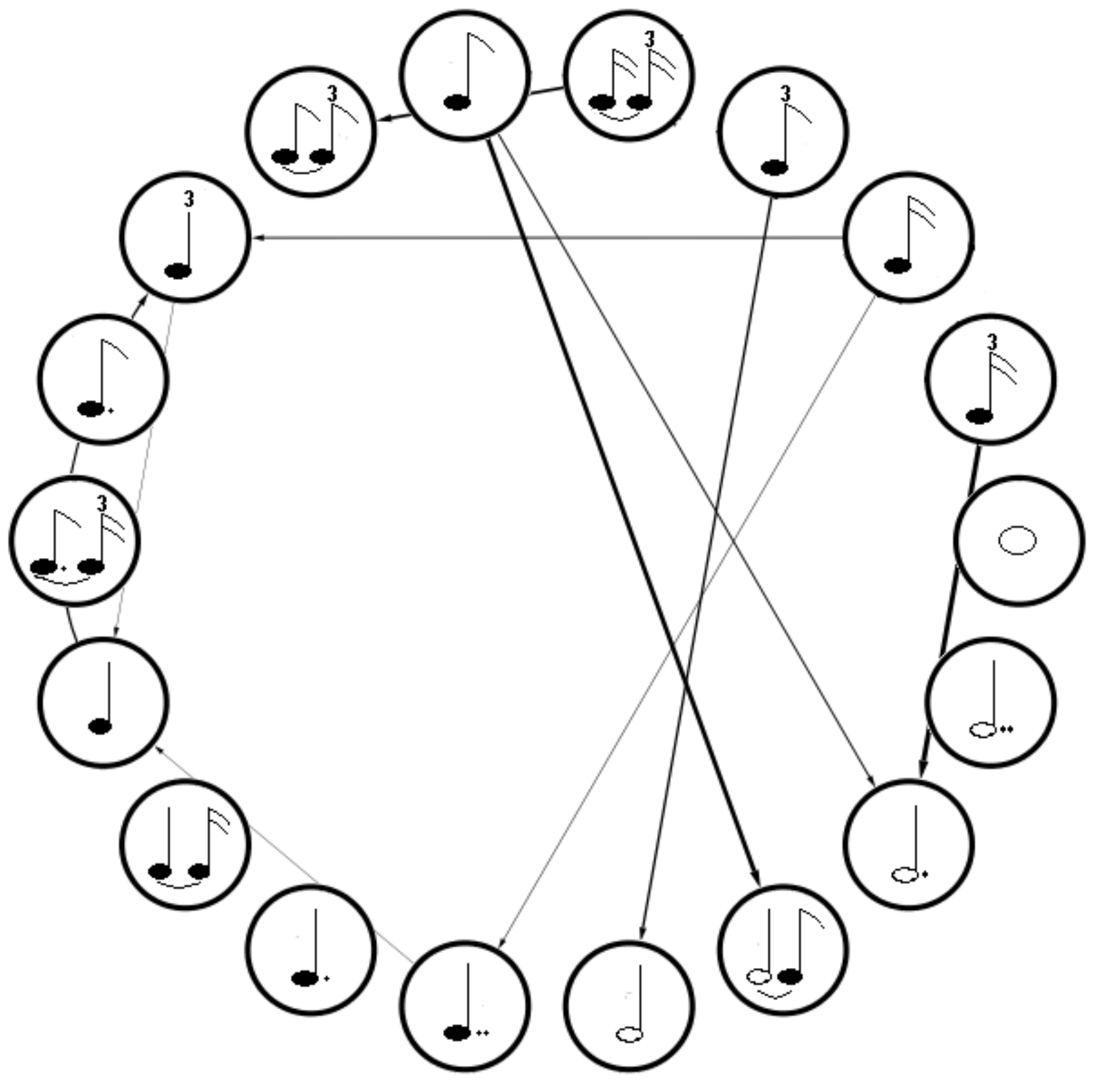}
   \label{fig:LDAAutoVetor1}
}
\hspace{0.3cm}
\subfigure[]
{
   \centering
   \includegraphics[width=7cm]{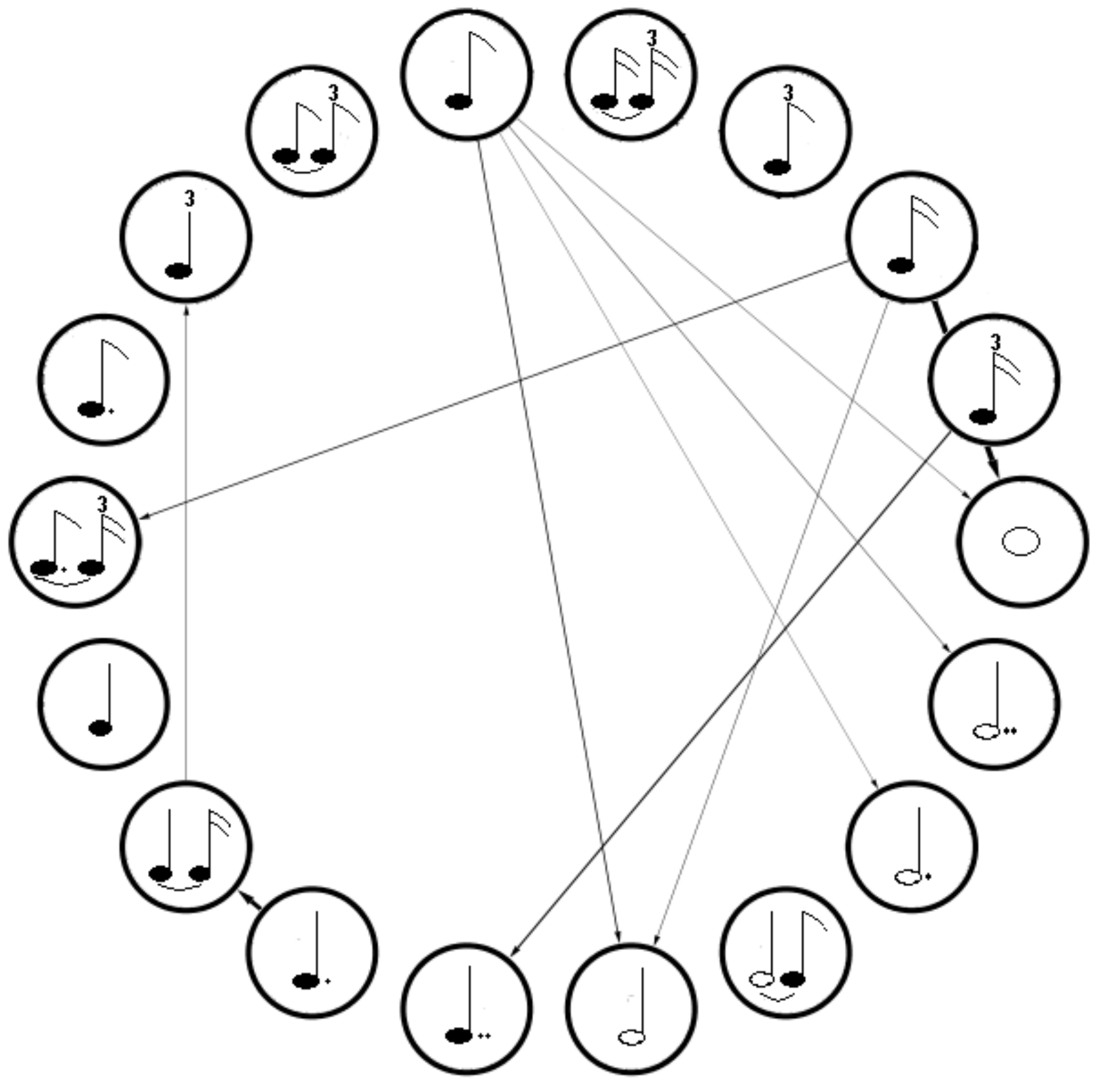}
   \label{fig:LDAAutoVetor2}
}
\caption{Ten most significant sequences of rhythm notations according to: (a) PCA first eigenvector (b) PCA second eigenvector (c) LDA first eigenvector (d) LDA second eigenvector}
\label{fig:autvetpcalda}
\end{figure}

The PCA and LDA approaches help to identify which features contribute
the most to the classification. This is an interesting analysis that
can be performed by verifying the strength of each element in the
first eigenvectors, and then associating those elements with the
original features. Within the current study, it was figured out that
the first ten sequences of rhythm notations that most contributed to
separation correspond to those illustrated in Figure
\ref{fig:autvetpcalda}. In the case of the first and second eigenvectors 
obtained by PCA and LDA, the ten elements with higher values were
selected, and the indices of these elements were associated with the
sequences in the original weight matrix. Figure
\ref{fig:PCAAutoVetor1} and \ref{fig:PCAAutoVetor2} shows the
resulting sequences according to the first and second eigenvectors of
PCA. The thickness of the edges is set by the value of the
corresponding element in the eigenvector. It is interesting that these
sequences are those that mostly frequently happen in the rhythms from
all four genres studied here. That is, they correspond to the elements
that play the greatest role in representing the rhythms. Therefore, it
can be said that these are the \textit{ten most representative
sequences}, when the first and second eigenvectors of PCA are taken
into account. Triples of eighth and sixteenth notes are particularly
important in blues and reggae genres. Similarly, Figure
\ref{fig:LDAAutoVetor1} and Figure
\ref{fig:LDAAutoVetor2} show the resulting sequences according to the
first and second eigenvectors of LDA. Differently from those obtained
by PCA, these sequences are not common to all the rhythms, but they
must happen with distinct frequency within each genre. Thus, they are
referred here as the \textit{ten most discriminative sequences}, when
the first and second eigenvectors of LDA are taken into account.

\clearpage

Clustering results are discussed in the following. The number of
clusters was defined as being four, in order to provide a fair
comparison with the supervised classification results. The idea of the
confusion matrix in Table \ref{tab:MatConfClustering} is to verify how
many art works from each class were placed in each one of the four
clusters. For example, it is known that art works from one to seventy
belongs to the blues genre. Then, the first line of this confusion
matrix indicates that four blues art works were placed in the cluster
one, thirty three in the cluster two, seven in the cluster three and
twenty six in the cluster four. It can also be observed that in
cluster one reggae art works are the majority (nineteen), despite the
small difference with the number of rock art works (fifteen); while in
cluster two the majority are blues art works; in cluster three the
majority are rock art works; and in cluster four the majority are
\textit{bossa nova} art works.

\begin{table}
    \centering
    \caption{Confusion Matrix for the Agglomerative Hierarchical Clustering}
    \footnotesize
        \begin{tabular}{l c c c c c}
        \hline
        &  Classe 1 & Classe 2 & Classe 3 & Classe 4  \\
        \hline\hline
        Classe 1      & 4  & 33  &  7 &  26 \\ 
        Classe 2      & 9  &  4  &  7 &  50 \\ 
        Classe 3      & 19 & 15  &  6 &  30 \\ 
        Classe 4      & 15 &  5  & 14 &  36 \\ 
        \hline\hline
        \end{tabular}
        \label{tab:MatConfClustering}
\end{table}

Comparing the confusion matrix in Table \ref{tab:MatConfClustering}
and the confusion matrix for the Quadradic Bayesian Classifier using
PCA in Table \ref{tab:MatConfPCA}, it is interesting to notice that:
in the former, the cluster four contains considerable art works from
the four genres (twenty six from blues, fifty from \textit{bossa
nova}, thirty from reggae and thirty six from rock), in a total of one
hundred forty two art works; in the later, a considerable number of
art work from blues (twenty two), \textit{bossa nova} (twenty one) and
rock (twenty four) were misclassified as reggae, in a total of one
hundred thirty seven art works belonging this class. This means that
the PCA representation was not efficient in discriminating reggae from
the other genres, while cluster four was the one that mostly
intermixed art works from all classes.

\begin{figure}
\centering
    \includegraphics[width=15cm]{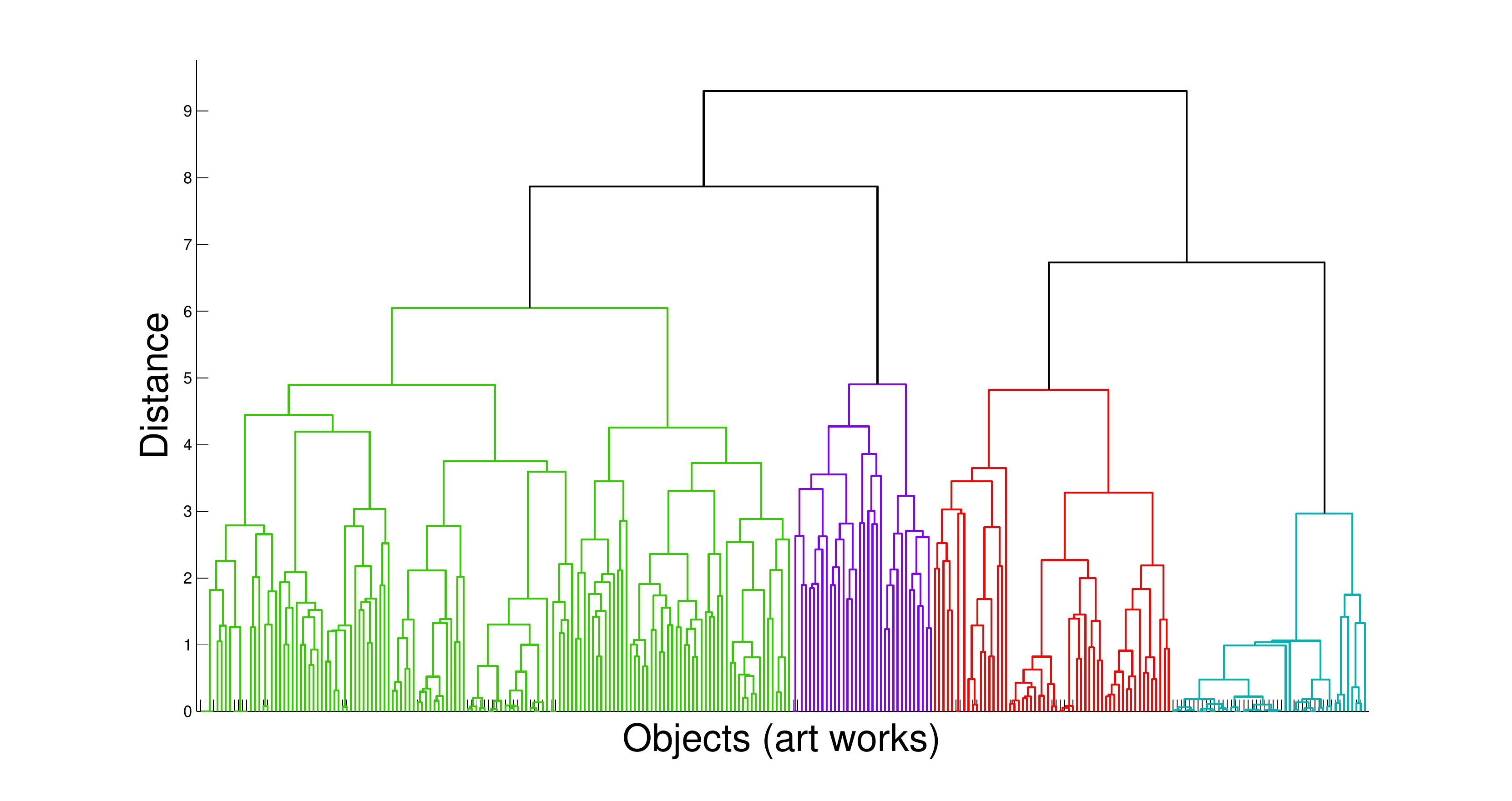}
  \caption{The dendrogram of the resulting four clusters (colored in green, pink, red and cyan.}
\label{fig:totaldendrogram}
\end{figure}

Figure \ref{fig:totaldendrogram} presents the dendrogram with the four
identified clusters. Different colors were used for the sake of
enhanced visual analysis. Cluster one is colored in green, cluster two
in pink, cluster three in red, and cluster four in cyan. These colors
were based on the dominant class in each cluster. For example, cluster
one is colored in green because \textit{bossa nova} art works are
majority in this cluster.

\begin{figure}
\centering
    \includegraphics[width=15cm]{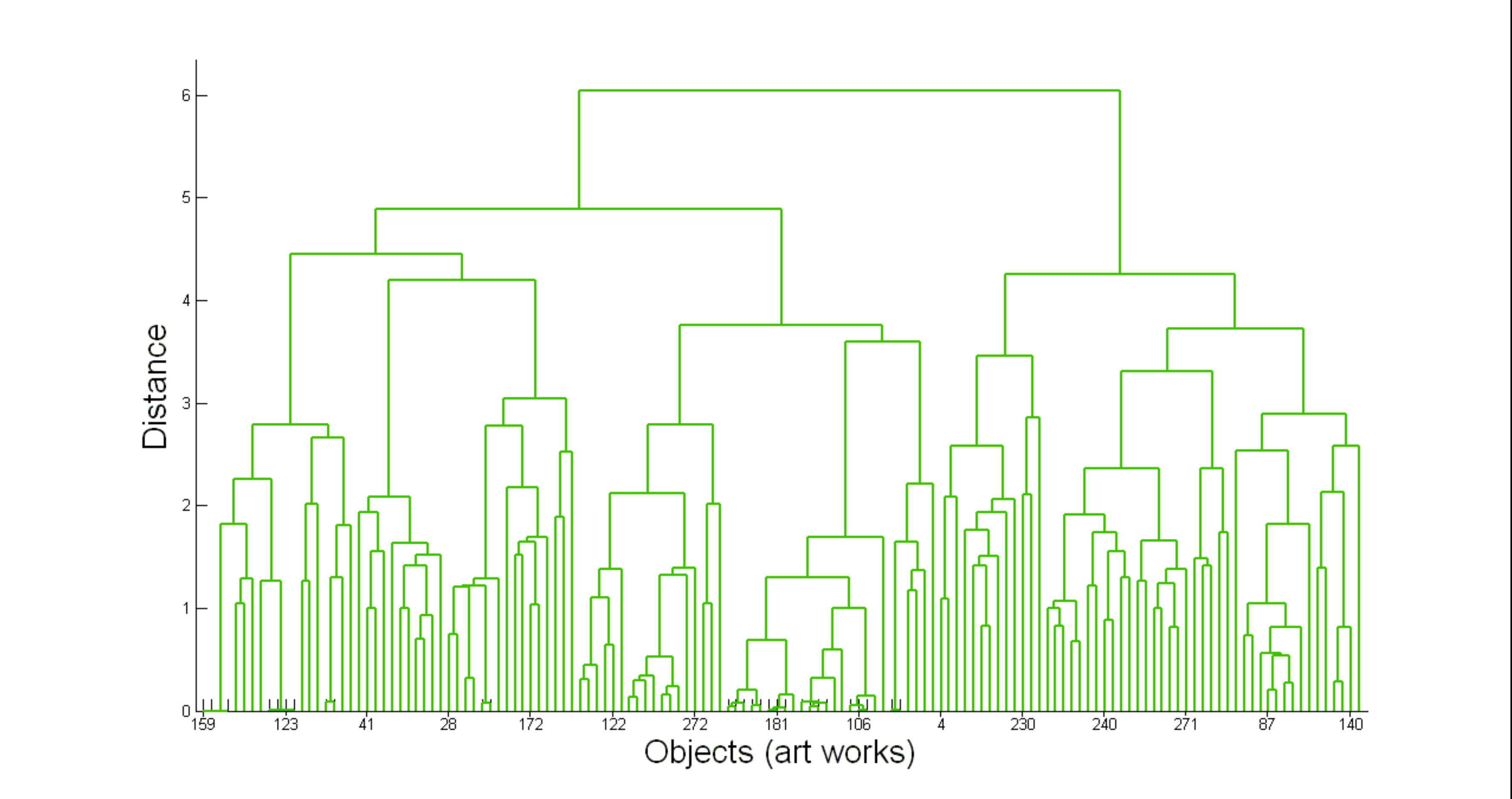}
  \caption{The detailed dendrogram for cluster in green. The objects were specified here, from left to right: \textbf{
  \textcolor[rgb]{0, 0.7, 0.7}{159}	\textcolor[rgb]{0.5, 0, 1}{237}	\textcolor[rgb]{0, 0.7, 0.7}{158}	\textcolor[rgb]{0.2, 0.8, 0}{74}	\textcolor{red}{37}	\textcolor[rgb]{0.2, 0.8, 0}{80}	\textcolor[rgb]{0.2, 0.8, 0}{132}	\textcolor[rgb]{0.2, 0.8, 0}{85}	\textcolor[rgb]{0.2, 0.8, 0}{117}	\textcolor[rgb]{0, 0.7, 0.7}{208} 	\textcolor[rgb]{0.2, 0.8, 0}{123}	\textcolor[rgb]{0.2, 0.8, 0}{126}	\textcolor{red}{2}	\textcolor[rgb]{0.2, 0.8, 0}{78}	\textcolor{red}{39}	\textcolor{red}{23} 	\textcolor{red}{62}	\textcolor{red}{45}	\textcolor[rgb]{0.2, 0.8, 0}{138}	\textcolor{red}{11}	\textcolor{red}{41}	\textcolor[rgb]{0, 0.7, 0.7}{156} 	\textcolor{red}{46}	\textcolor[rgb]{0.2, 0.8, 0}{88}	\textcolor[rgb]{0.2, 0.8, 0}{114}	\textcolor[rgb]{0.2, 0.8, 0}{131}	\textcolor[rgb]{0, 0.7, 0.7}{164}	\textcolor[rgb]{0.5, 0, 1}{266}	\textcolor[rgb]{0, 0.7, 0.7}{166}	\textcolor[rgb]{0.5, 0, 1}{258}	\textcolor{red}{28} \textcolor[rgb]{0, 0.7, 0.7}{174}	\textcolor[rgb]{0.2, 0.8, 0}{105}	\textcolor[rgb]{0.5, 0, 1}{233}	\textcolor[rgb]{0, 0.7, 0.7}{141}	\textcolor[rgb]{0, 0.7, 0.7}{167}	\textcolor[rgb]{0, 0.7, 0.7}{184}	\textcolor[rgb]{0.2, 0.8, 0}{81}	\textcolor[rgb]{0, 0.7, 0.7}{163}	\textcolor[rgb]{0, 0.7, 0.7}{207}	\textcolor[rgb]{0, 0.7, 0.7}{172}	\textcolor[rgb]{0, 0.7, 0.7}{182}	\textcolor[rgb]{0.5, 0, 1}{248}	\textcolor[rgb]{0.2, 0.8, 0}{72}	\textcolor[rgb]{0.2, 0.8, 0}{95}	\textcolor[rgb]{0.5, 0, 1}{216}	\textcolor{red}{13}	\textcolor[rgb]{0, 0.7, 0.7}{185}	\textcolor[rgb]{0.2, 0.8, 0}{96}	\textcolor[rgb]{0.2, 0.8, 0}{112}	\textcolor[rgb]{0.2, 0.8, 0}{122}	\textcolor[rgb]{0.2, 0.8, 0}{111}	\textcolor[rgb]{0.2, 0.8, 0}{98}	\textcolor[rgb]{0, 0.7, 0.7}{175}	\textcolor[rgb]{0, 0.7, 0.7}{209}	\textcolor[rgb]{0.5, 0, 1}{266}	\textcolor[rgb]{0, 0.7, 0.7}{177}	 \textcolor[rgb]{0.5, 0, 1}{247}	\textcolor[rgb]{0.5, 0, 1}{212}	\textcolor[rgb]{0, 0.7, 0.7}{206}	\textcolor[rgb]{0.5, 0, 1}{272}	\textcolor[rgb]{0, 0.7, 0.7}{179}	\textcolor[rgb]{0.5, 0, 1}{225}	\textcolor[rgb]{0.5, 0, 1}{253}	\textcolor{red}{52}	\textcolor{red}{54}	\textcolor{red}{63}	\textcolor[rgb]{0.2, 0.8, 0}{89}	\textcolor[rgb]{0.5, 0, 1}{262}	\textcolor[rgb]{0.2, 0.8, 0}{86}	\textcolor[rgb]{0, 0.7, 0.7}{181}	\textcolor[rgb]{0.5, 0, 1}{254}	\textcolor[rgb]{0.5, 0, 1}{276}	\textcolor{red}{58}	\textcolor[rgb]{0.2, 0.8, 0}{118}	\textcolor[rgb]{0.5, 0, 1}{264}	\textcolor[rgb]{0, 0.7, 0.7}{197}	\textcolor[rgb]{0.2, 0.8, 0}{115}	\textcolor[rgb]{0.2, 0.8, 0}{116}	\textcolor[rgb]{0.2, 0.8, 0}{82}	\textcolor[rgb]{0.2, 0.8, 0}{106}	\textcolor[rgb]{0.2, 0.8, 0}{125}	\textcolor[rgb]{0, 0.7, 0.7}{169}	\textcolor[rgb]{0.5, 0, 1}{220}	\textcolor[rgb]{0.2, 0.8, 0}{119}	\textcolor[rgb]{0.2, 0.8, 0}{134}	\textcolor[rgb]{0, 0.7, 0.7}{162}	\textcolor[rgb]{0.5, 0, 1}{243}	\textcolor[rgb]{0.5, 0, 1}{250}	\textcolor[rgb]{0, 0.7, 0.7}{201}	\textcolor{red}{4}	\textcolor{red}{15}	\textcolor[rgb]{0, 0.7, 0.7}{199}	\textcolor{red}{14}	\textcolor{red}{53}	\textcolor{red}{57}	\textcolor[rgb]{0.2, 0.8, 0}{93}	\textcolor[rgb]{0.5, 0, 1}{244}	\textcolor[rgb]{0.5, 0, 1}{242}	\textcolor[rgb]{0, 0.7, 0.7}{165}	\textcolor[rgb]{0.5, 0, 1}{230}	\textcolor[rgb]{0.5, 0, 1}{265}	\textcolor[rgb]{0.5, 0, 1}{232}	\textcolor{red}{19}	\textcolor{red}{55}	\textcolor[rgb]{0.2, 0.8, 0}{136}	\textcolor[rgb]{0.2, 0.8, 0}{104}	\textcolor[rgb]{0.5, 0, 1}{238}	\textcolor[rgb]{0.5, 0, 1}{215}	\textcolor[rgb]{0.5, 0, 1}{251}	\textcolor[rgb]{0.5, 0, 1}{240}	\textcolor[rgb]{0.5, 0, 1}{252}	\textcolor[rgb]{0.5, 0, 1}{241}	\textcolor[rgb]{0.5, 0, 1}{269}	\textcolor{red}{60}	\textcolor[rgb]{0, 0.7, 0.7}{195}	\textcolor{red}{68}	\textcolor[rgb]{0, 0.7, 0.7}{148}	\textcolor[rgb]{0.5, 0, 1}{277}	\textcolor[rgb]{0.5, 0, 1}{280}	\textcolor[rgb]{0.5, 0, 1}{271}	\textcolor[rgb]{0.2, 0.8, 0}{101}	\textcolor[rgb]{0.2, 0.8, 0}{110}	\textcolor[rgb]{0.5, 0, 1}{268}	\textcolor[rgb]{0.2, 0.8, 0}{137}	\textcolor[rgb]{0, 0.7, 0.7}{186}	\textcolor{red}{22}	\textcolor{red}{30}	\textcolor[rgb]{0.2, 0.8, 0}{71}	\textcolor[rgb]{0.2, 0.8, 0}{73}	\textcolor[rgb]{0.2, 0.8, 0}{87}	\textcolor[rgb]{0.2, 0.8, 0}{108}	\textcolor[rgb]{0.2, 0.8, 0}{121}	\textcolor[rgb]{0.2, 0.8, 0}{130}	\textcolor[rgb]{0.2, 0.8, 0}{129}	\textcolor[rgb]{0.2, 0.8, 0}{133}	\textcolor[rgb]{0.2, 0.8, 0}{113}	\textcolor[rgb]{0.2, 0.8, 0}{135}	\textcolor[rgb]{0.2, 0.8, 0}{124}	\textcolor[rgb]{0.2, 0.8, 0}{127}	\textcolor[rgb]{0.2, 0.8, 0}{140}	and \textcolor[rgb]{0.5, 0, 1}{221}. }}
\label{fig:dendrogramcluster4}
\end{figure}

\begin{figure}
\centering
    \includegraphics[width=15cm]{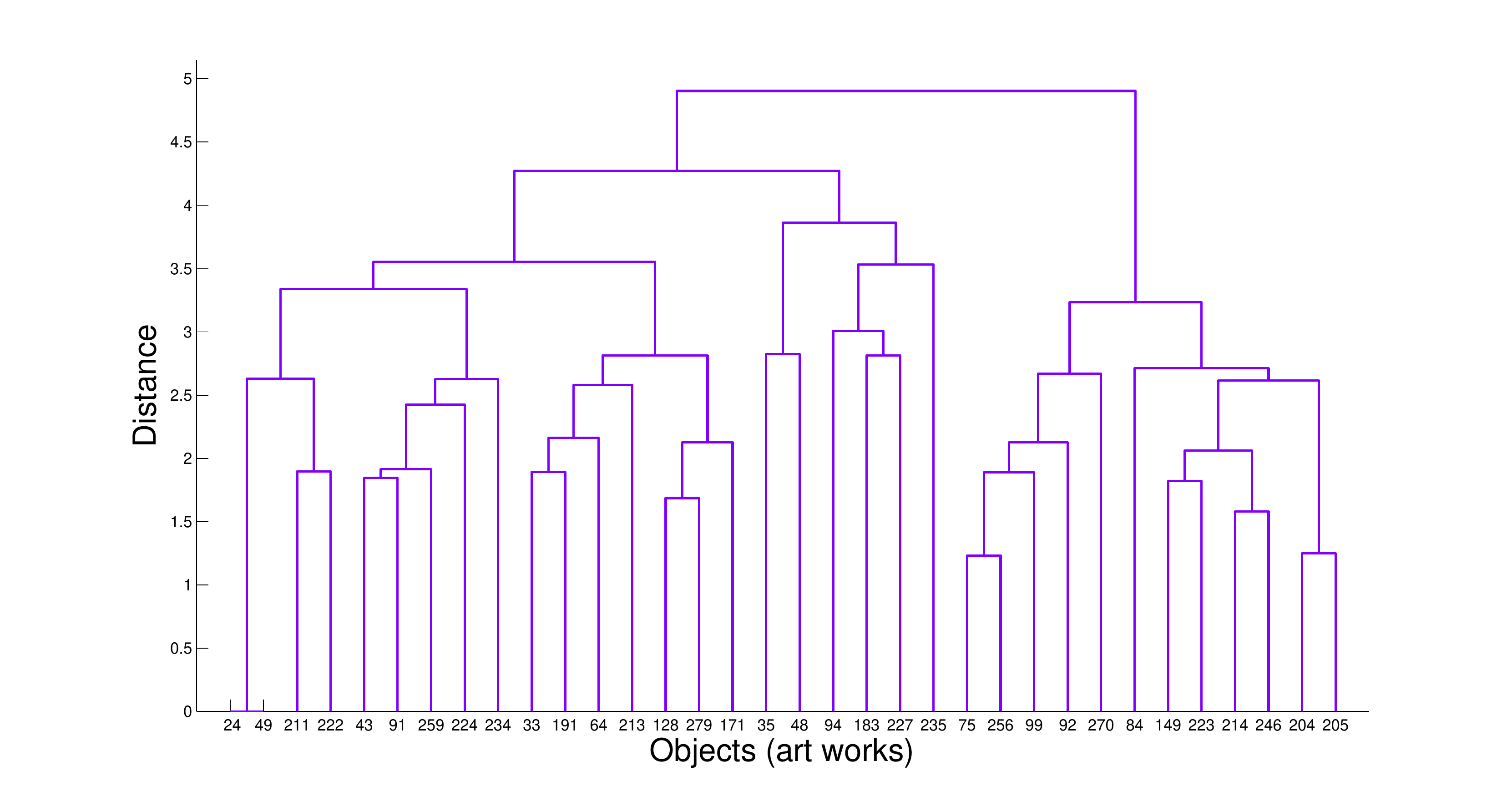}
  \caption{The detailed dendrogram for cluster in pink.The objects were specified here, from left to right: \textbf{\textcolor{red}{24} \textcolor{red}{49} \textcolor[rgb]{0.5, 0, 1}{211}   \textcolor[rgb]{0.5, 0, 1}{222}    \textcolor{red}{43} \textcolor[rgb]{0.2, 0.8, 0}{91}   \textcolor[rgb]{0.5, 0, 1}{259}   \textcolor[rgb]{0.5, 0, 1}{224}   \textcolor[rgb]{0.5, 0, 1}{234}    \textcolor{red}{33}   \textcolor[rgb]{0,0.7,0.7}{191}    \textcolor{red}{64}   \textcolor[rgb]{0.5, 0, 1}{213}   \textcolor[rgb]{0.2, 0.8, 0}{128}   \textcolor[rgb]{0.5, 0, 1}{279}   \textcolor[rgb]{0,0.7,0.7}{171}    \textcolor{red}{35} \textcolor{red}{48} \textcolor[rgb]{0.2, 0.8, 0}{94}   \textcolor[rgb]{0,0.7,0.7}{183} \textcolor[rgb]{0.5, 0, 1}{227}   \textcolor[rgb]{0.5, 0, 1}{235}    \textcolor[rgb]{0.2, 0.8, 0}{75}   \textcolor[rgb]{0.5, 0, 1}{256}    \textcolor[rgb]{0.2, 0.8, 0}{99}    \textcolor[rgb]{0.2, 0.8, 0}{92}   \textcolor[rgb]{0.5, 0, 1}{270}    \textcolor[rgb]{0.2, 0.8, 0}{84}   \textcolor[rgb]{0,0.7,0.7}{149}   \textcolor[rgb]{0.5, 0, 1}{223}   \textcolor[rgb]{0.5, 0, 1}{214}   \textcolor[rgb]{0.5, 0, 1}{246}   \textcolor[rgb]{0,0.7,0.7}{204} and \textcolor[rgb]{0,0.7,0.7}{205}}}.
\label{fig:dendrogramcluster3}
\end{figure}

\begin{figure}
\centering
    \includegraphics[width=15cm]{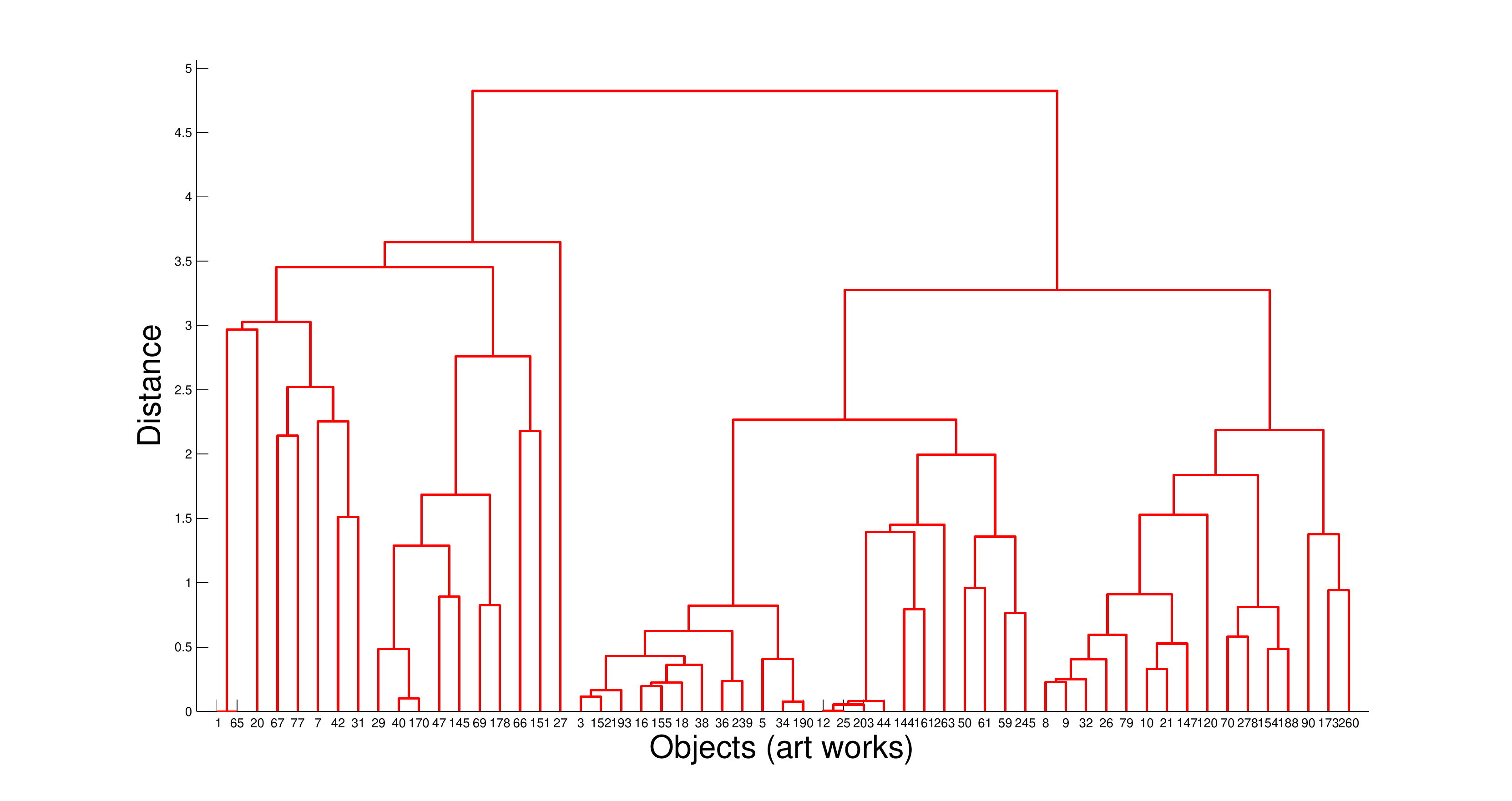}
  \caption{The detailed dendrogram for cluster in red.The objects were specified here, from left to right: \textbf{ \textcolor{red}{1}    \textcolor{red}{65}    \textcolor{red}{20}    \textcolor{red}{67}    \textcolor[rgb]{0.2,0.8,0}{77}     \textcolor{red}{7}    \textcolor{red}{42}    \textcolor{red}{31}    \textcolor{red}{29}    \textcolor{red}{40}   \textcolor[rgb]{0,0.7,0.7}{170}    \textcolor{red}{47}   \textcolor[rgb]{0,0.7,0.7}{145}    \textcolor{red}{69}   \textcolor[rgb]{0,0.7,0.7}{178}    \textcolor{red}{66}   \textcolor[rgb]{0,0.7,0.7}{151}  \textcolor{red}{27}  \textcolor{red}{3}   \textcolor[rgb]{0,0.7,0.7}{152}   \textcolor[rgb]{0,0.7,0.7}{193}    \textcolor{red}{16}   \textcolor[rgb]{0,0.7,0.7}{155}    \textcolor{red}{18}    \textcolor{red}{38}    \textcolor{red}{36}   \textcolor[rgb]{0.5, 0, 1}{239}     \textcolor{red}{5}    \textcolor{red}{34}   \textcolor[rgb]{0,0.7,0.7}{190}    \textcolor{red}{12}    \textcolor{red}{25}   \textcolor[rgb]{0,0.7,0.7}{203} \textcolor{red}{44}  \textcolor[rgb]{0,0.7,0.7}{144}   \textcolor[rgb]{0,0.7,0.7}{161}   \textcolor[rgb]{0.5, 0, 1}{263}    \textcolor{red}{50}    \textcolor{red}{61}    \textcolor{red}{59}   \textcolor[rgb]{0.5, 0, 1}{245}     \textcolor{red}{8}     \textcolor{red}{9}    \textcolor{red}{32}    \textcolor{red}{26}    \textcolor[rgb]{0.2,0.8,0}{79}    \textcolor{red}{10}    \textcolor{red}{21}   \textcolor[rgb]{0,0.7,0.7}{147}   \textcolor[rgb]{0.2,0.8,0}{120}    \textcolor{red}{70} \textcolor[rgb]{0.5, 0, 1}{278}   \textcolor[rgb]{0,0.7,0.7}{154}   \textcolor[rgb]{0,0.7,0.7}{188}    \textcolor[rgb]{0.2,0.8,0}{90}   \textcolor[rgb]{0,0.7,0.7}{173}  and \textcolor[rgb]{0.5, 0, 1}{260}.}}
\label{fig:dendrogramcluster2}
\end{figure}

\begin{figure}
\centering
    \includegraphics[width=15cm]{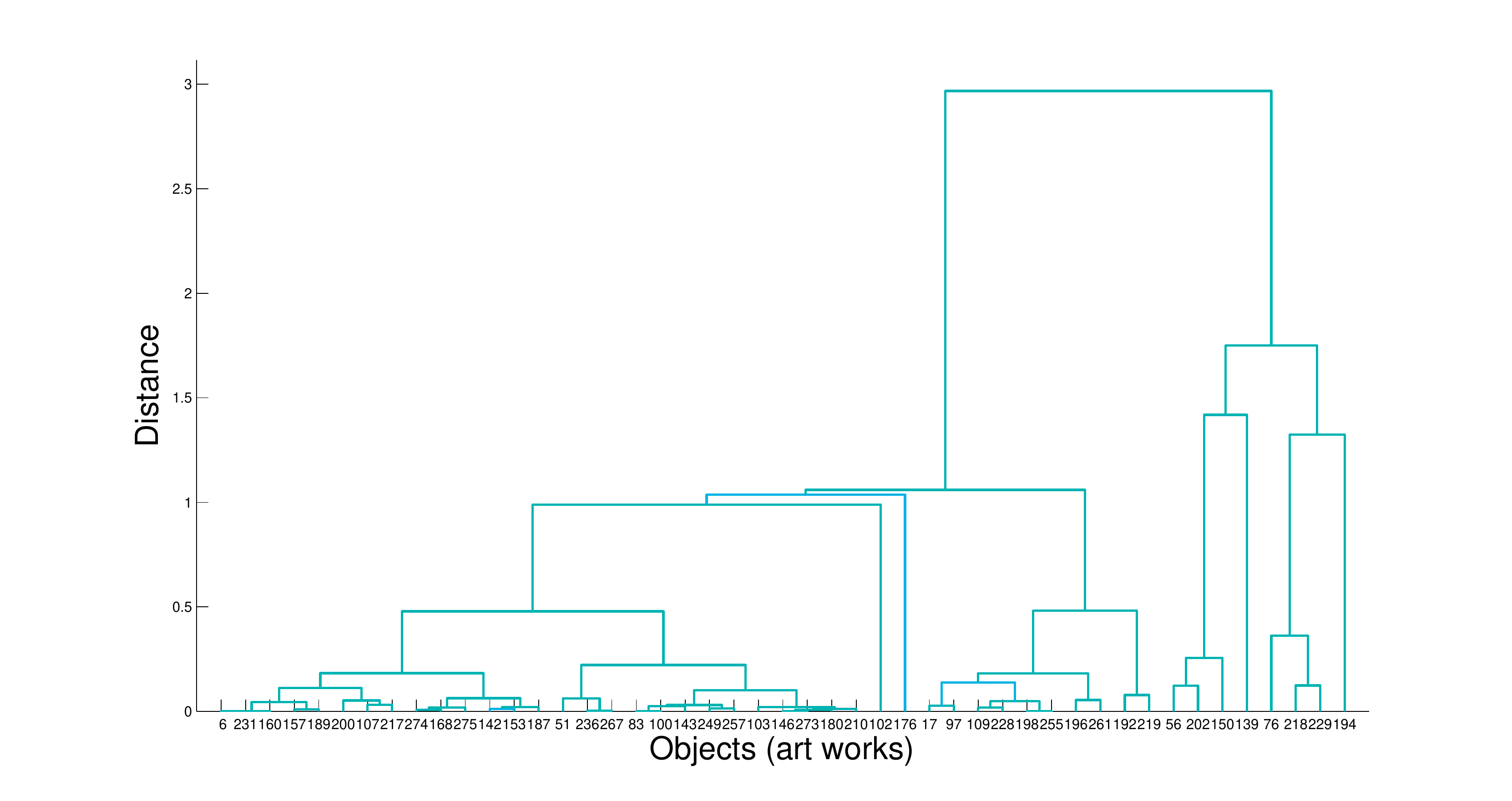}
  \caption{The detailed dendrogram for cluster in cyan. The objects were specified here, from left to right: \textbf{ \textcolor{red}{6}   \textcolor[rgb]{0.5, 0, 1}{231}   \textcolor[rgb]{0,0.7,0.7}{160}   \textcolor[rgb]{0,0.7,0.7}{157}   \textcolor[rgb]{0,0.7,0.7}{189}   \textcolor[rgb]{0,0.7,0.7}{200}   \textcolor[rgb]{0.2, 0.8, 0}{107}   \textcolor[rgb]{0.5, 0, 1}{217}   \textcolor[rgb]{0.5, 0, 1}{274}   \textcolor[rgb]{0,0.7,0.7}{168}   \textcolor[rgb]{0.5, 0, 1}{275}   \textcolor[rgb]{0,0.7,0.7}{142}   \textcolor[rgb]{0,0.7,0.7}{153}   \textcolor[rgb]{0,0.7,0.7}{187}    \textcolor{red}{51}   \textcolor[rgb]{0.5, 0, 1}{236}   \textcolor[rgb]{0.5, 0, 1}{267} \textcolor[rgb]{0.2, 0.8, 0}{83}   \textcolor[rgb]{0.2, 0.8, 0}{100}   \textcolor[rgb]{0,0.7,0.7}{143}   \textcolor[rgb]{0.5, 0, 1}{249}   \textcolor[rgb]{0.5, 0, 1}{257}   \textcolor[rgb]{0.2, 0.8, 0}{103}   \textcolor[rgb]{0,0.7,0.7}{146}   \textcolor[rgb]{0.5, 0, 1}{273}   \textcolor[rgb]{0,0.7,0.7}{180}   \textcolor[rgb]{0,0.7,0.7}{210}   \textcolor[rgb]{0.2, 0.8, 0}{102}   \textcolor[rgb]{0,0.7,0.7}{176}    \textcolor{red}{17}    \textcolor[rgb]{0.2, 0.8, 0}{97}   \textcolor[rgb]{0.2, 0.8, 0}{109}   \textcolor[rgb]{0.5, 0, 1}{228}   \textcolor[rgb]{0,0.7,0.7}{198}  \textcolor[rgb]{0.5, 0, 1}{255}   \textcolor[rgb]{0,0.7,0.7}{196}   \textcolor[rgb]{0.5, 0, 1}{261}   \textcolor[rgb]{0,0.7,0.7}{192}   \textcolor[rgb]{0.5, 0, 1}{219}    \textcolor{red}{56}   \textcolor[rgb]{0,0.7,0.7}{202}   \textcolor[rgb]{0,0.7,0.7}{150}   \textcolor[rgb]{0.2, 0.8, 0}{139}    \textcolor[rgb]{0.2, 0.8, 0}{76}   \textcolor[rgb]{0.5, 0, 1}{218}   \textcolor[rgb]{0.5, 0, 1}{229}  and \textcolor[rgb]{0,0.7,0.7}{194}. }}
\label{fig:dendrogramcluster1}
\end{figure}

The four obtained clusters are detailed in Figures
\ref{fig:dendrogramcluster4} to \ref{fig:dendrogramcluster1}, in which
the legends presents the grouped art works from each cluster (blues
art works are in red, \textit{bossa nova} art works are in green,
reggae art works in cyan and rock art works in pink).

As a consequence of working in a higher dimension feature space, the
agglomerative hierarquical clustering approach could better separate
the data when compared to the PCA and LDA-based approaches, which are
applied over a projected version of the original measurements.

\section{Concluding Remarks}\label{sec:Conclusions}

Automatic music genre classification has become a fundamental topic in
music research since genres have been widely used to organize and
describe music collections. They also reveal general identities of
different cultures. However, music genres are not a clearly defined
concept so that the development of a non-controversial taxonomy
represents a challenging, non-trivial taks.

Generally speaking, music genres summarize common characteristics of
musical pieces. This is particular interesting when it is used as a
resource for automatic classification of pieces. In the current paper,
we explored genre classification while taking into account the musical
temporal aspects, namely the rhythm. We considered pieces of four
musical genres (blues, \textit{bossa nova}, reggae and rock), which
were extracted from MIDI files and modeled as networks. Each node
corresponded to one rhythmc notation, and the links were defined by
the sequence in which they occurred along time. The idea of using
static nodes (nodes with fixed positions) is particularly interesting
because it provides a primary visual identification of the differences
and similarities between the rythms from the four genres. A Markov
model was build from the networks, and the dynamics and dependencies
of the rhythmic notations were estimated, comprising the feature
matrix of the data. Two different approaches for features analysis
were used (PCA and LDA), as well as two types of classification
methods (Bayesian classifier and hierarchical clustering).

Using only the first two principal componentes, the different types of
rhythms were not separable, although for the first and third axes we
could observe some separation between three of the classes (blues,
\textit{bossa nova} and reggae), while only the samples of rock
overlapped the other classes.  However, taking into account that
twenty components were necessary to preserve 76\% of the data
variance, it would be expected that only two or three dimensions would
not be sufficient to allow suitable saparability. Notably, the
dimensionality of the problem is high, that is, the rhythms are very
complex and many dimensions (features) are necessary to separate them.
This is one of the main findings of the current work.  With the help
of LDA analysis, another finding was reached which supported the
assumption that the problem of automatic rhythm classification is no
trivial task.  The projections obtained by considering the first and
second, and first and third axes implied in better discrimination
between the four classes than that obtained by the PCA.  

Unlike PCA and LDA, agglomerative hierarchical clustering works on the
original dimensions of the data.  The application of the methodology
led to a substantially better discrimination, which provides a strong
evidence of the complexity of the problem studied here. The results
are promising in the sense that in each cluster is dominated by a
different genre, showing the viability of the proposed approach. 

It is clear from our study that musical genres are very complex and
present redundancies. Sometimes it is difficult even for an expert to
distinguish them. This difficulty becomes more critical when only the
rhythm is taken into account.

Several are the possibilities for future research implied by the
reported investigation.  First, it would be interesting to use more
measurements extracted from rhythm, especially the intensity of the
beats, as well as the distribution of instruments, which is poised to
improve the classification results.  Another promising venue for
further investigation regards the use of other classifiers, as well as
the combination of results obtained from ensemble of distinct
classifiers.  In addition, it would be promising to apply
multi-labeled classification, a growing field of research in which
non-disjointed samples can be associated to one or more labels
\cite{TSOUMAKAS2007}. Nowadays, multi-labeled classification methods
have been increasingly required by applications such as text
categorization \cite{KATAKIS2008}, scene classification
\cite{BOUTELL2004}, protein classification \cite{DIPLARIS2005}, and
music categorization in terms of emotion \cite{TROHIDIS2008}, among
others. The possibility of multigenres classification is particularly
promising and probably closer to the human experience.  Another
interesting future work is related to the synthesis of rhythms. Once
the rhythmic networks are available, new rhythms with similar
characteristics according to the specific genre can be artificially
generated.


\section{Acknowledgments}\label{sec:Acknowledgments}
Debora C Correa is grateful to FAPESP (2009/50142-0) for financial
support and Luciano da F. Costa is grateful to CNPq (301303/06-1 and
573583/2008-0) and FAPESP (05/00587-5) for financial support.

\clearpage
\section*{Appendices}

\appendix
\section{Multivariate Statistical Methods} \label{sec:APCALDA}

\subsection{Principal Component Analysis} \label{subsec:PCA}

Principal Component Analysis is a second order unsupervised
statistical technique. By second order it is meant that all the
necessary information is available directly from the covariance matrix
of the mixture data, so that there is no need to use the complete
probability distributions. This method uses the eigenvalues and
eigenvectors of the covariance matrix in order to transform the
feature space, creating orthogonal uncorrelated features. From a
multivariate dataset, the principal aim of PCA is to remove redundancy
from the data, consequently reducing the dimensionality of the
data. Additional information about PCA and its relation to various
interesting statistical and geometrical properties can be found in the
pattern recognition literature, e.g. \cite{HYVARINEN2001, WEBB2002,
DUDA2001, COSTA2001}.

Consider a vector \textbf{x} with \textit{n} elements representing
some features or measurements of a sample. In the first step of PCA
transform, this vector \textbf{x} is centered by subtracting its mean,
so that $ \textbf{x} \leftarrow \textbf{x} - E
\left\{\textbf{x}\right\}$. Next, \textbf{x} is linearly transformed
to a different vector \textbf{y} which contains \textit{m} elements,
\textit{m} $ < $ \textit{n}, removing the redundancy caused by the
correlations. This is achieved by using a rotated orthogonal
coordinate system in such a way that the elements in \textbf{x} are
uncorrelated in the new coordinate system. At the same time, PCA
maximizes the variances of the projections of \textbf{x} on the new
coordinate axes (components). These variances of the components will
differ in most applications.  The axes associated to small dispersions
(given by the respectively associated eigenvalues) can be discarded
without losing too much information about the original data.
 
\subsection{Linear discriminant Analysis} \label{subsec:LDA}

Linear discriminantAnalysis (LDA) can be considered a generalization
of Fisher's Linear discriminant Function for the multivariate case
\cite{DUDA2001, WEBB2002}.  It is a supervised approach that maximizes
data separability, in terms of a simililarity criterion based on
scatter matrices. The basic idea is that objects belonging to the same
class are as similar as possible and objects belonging to distinct
classes are as different as possible. In other words, LDA looks for a
new, projected, feature space where that maximizes interclass distance
while minimizing the intraclass distance. This result can be later
used for linear classification, and it is also possible to reduce
dimensionality before the classification task.  The scatter matrix for
each class indicates the dispersion of the features vectors within the
class. The intraclass scatter matrix is defined as the sum of the
scatter matrices of all classes and expresses the combined dispersion
in each class. The interclass scatter matrix quantifies how disperse
the classes are, in terms of the position of their centroids.

It can be shown that the maximization criterion for class separability
leads to a generalized eigenvalue problem (\cite{WEBB2002,
DUDA2001}). Therefore, it is possible to compute the eigenvalues and
eigenvectors of the matrix defined by
$\left(S_{intra}^{-1}*S_{inter}\right)$, where $S_{intra}$ is the
intraclass scatter matrix and $S_{inter}$ is the interclass scatter
matrix. The \textit{m} eigenvectors associated to the \textit{m}
largest eigenvalues of this matrix can be used to project the
data. However, the rank of $\left(S_{intra}^{-1}*S_{inter}\right)$ is
limited to $\textit{C}-1$, where $\textit{C}$ is the number of
classes. As a consequence, there are $\textit{C}-1$ nonzero
eigenvalues, that is, the number of new features is conditioned to the
number of classes, $\textit{m} \leq
\textit{C}-1$. Another issue is that, for high dimensional problems,
when the number of available training samples is smaller than the
number of features, $S_{intra}$ becomes singular, complicating the
generalized eigenvalue solution.

More information about the LDA is \cite{DEVIJVER1981, Therrien1989, 
DUDA2001, WEBB2002}.

\clearpage
\section{Linear and Quadratic Discriminant Functions} \label{sec:ALinearDiscFunc}

If normal distribution over the data is assumed, it is possible to state
that:

\begin{equation}
p\left( x | \omega_i \right) = \frac{1}{\left( 2\pi \right)^{\frac{d}{2}} \left| \Sigma \right|^{\frac{1}{2}}} \exp \left\{ - \frac{1}{2} \left( \vec{x} - \vec{\mu} \right)^T \Sigma^{-1} \left( \vec{x} - \vec{\mu} \right) \right\}
\end{equation}

The components of the parameter vector for class $j$, $\vec{\theta}_j = \left\{ \vec{\mu}_j, \Sigma_j \right \}$, where $\vec{\mu}_j$ and $\Sigma _j $ are the mean vector
and the covariance matrix of class $j$, respectively, can be estimated
by maximum likelihood as follows:

\begin{equation}
\hat{\vec{\mu}}_j = \frac{1}{N} \sum_{i=1}^{N} \vec{x}_i
\end{equation}

\begin{equation}
\hat{\vec{\Sigma}}_j = \frac{1}{N} \sum_{i=1}^N \left( \vec{x}_i - \vec{\mu}_i \right) \left( \vec{x}_i - \vec{\mu}_i \right)^T
\end{equation}

Within this context, classification can be achieved with discriminant
functions, $g_i$, assigning an observed pattern vector $\vec{x}_i$ to the class $\omega_j$ with the maximum discriminant
function value.  By using Bayes's rule, not considering the constant
terms, and using the estimated parameters above, a decision rule can
be defined as: assign an object $\vec{x}_i$ to class
$\omega_j$ if $g_j > g_i $ for all $i \ne j$, where the
discriminantfunction $g_i$ is calculated as:

\begin{equation}
g_i \left( \vec{x} \right) = \log \left( p \left( \omega_i \right) \right) - \frac{1}{2} \log \left( \left| \hat{\Sigma}_i \right|\right) - \frac{1}{2} \left( \vec{x} - \hat{\vec{\mu}}_j \right)^T \hat{\Sigma}_i^{-1} \left( \vec{x} - \hat{\vec{\mu}}_j \right) 
\end{equation}

Classifying an object or pattern $\vec{x}$ on the basis
of the values of $g_i \left( \vec{x} \right)$, $i =
1,...,C$ ($C$ is the number of classes), with estimated parameters,
defines a quadratic discriminant classifier or quadratic Bayesian
classifier or yet quadratic Gaussian classifier \cite{WEBB2002}.

The prior probability, $p\left( {\omega _i } \right)$, can be simply
estimated by:

\begin{equation}
p\left( {\omega _i } \right) = \frac{{n_i }}{{\sum\nolimits_j {n_j } }}
\end{equation}

where $n_i$ is the number of samples of class $\omega_i$.

In multivariate classification situations, with different covariance
matrices, problems may occur in the quadratic Bayesian classifier when
any of the matrices $\hat{\Sigma}_i$ is
singular. This usually happens when there are not enough data to
obtain efficient estimative for the covariance matrices $ \Sigma _i $,
$i = 1,2,...,C $. An alternative to minimize this problem consist of
estimating one unique covariance matrix over all classes, $ \hat{\Sigma} =
\hat{\Sigma}_1 = ... = \hat{\Sigma}_C $. In this case, the discriminantfunction
becomes linear in $ \vec{x} $ and can be simplified:

\begin{equation}
	g_{i}\left( \vec{x} \right) = log \left( p \left( \omega_{i} \right) \right) - \frac{1}{2} \hat{\vec{\mu}}_{i}^{T} \hat{\Sigma}^{-1} \hat{\vec{\mu}}_{i} + \vec{x}^{T} \hat{\Sigma}^{-1} \hat{\vec{\mu}}_{i}
\end{equation}

where $\hat{\Sigma}$ is the covariance matrix, common to all classes.  The
classification rule remains the same. This defines a linear
discriminant classifier (also known as linear Bayesian classifier or
linear Gaussian classifier) \cite{WEBB2002}.

\clearpage 
\section{Agglomerative Hierarchical Clustering} \label{sec:HierClust}

Agglomerative hierarchical clustering groups progressively the $N$
objects into $C$ classes according to a defined parameter. The
distance or similarity between the feature vectors of the objects are
usually taken as such parameter. In the first step, there is a
partition with $N$ clusters, each cluster containing one object. The
next step is a different partition, with $N-1$ clusters, the next a
partition with $N-2$ clusters, and so on. In the $n$th step, all the
objects form a unique cluster. This sequence groups objects that are
more similar to one another into subclasses before objects that are
less similar. It is possible to say that, in the $k$th step, $C = N -
k + 1$.

\begin{figure}
\centering
    \includegraphics[width=8cm]{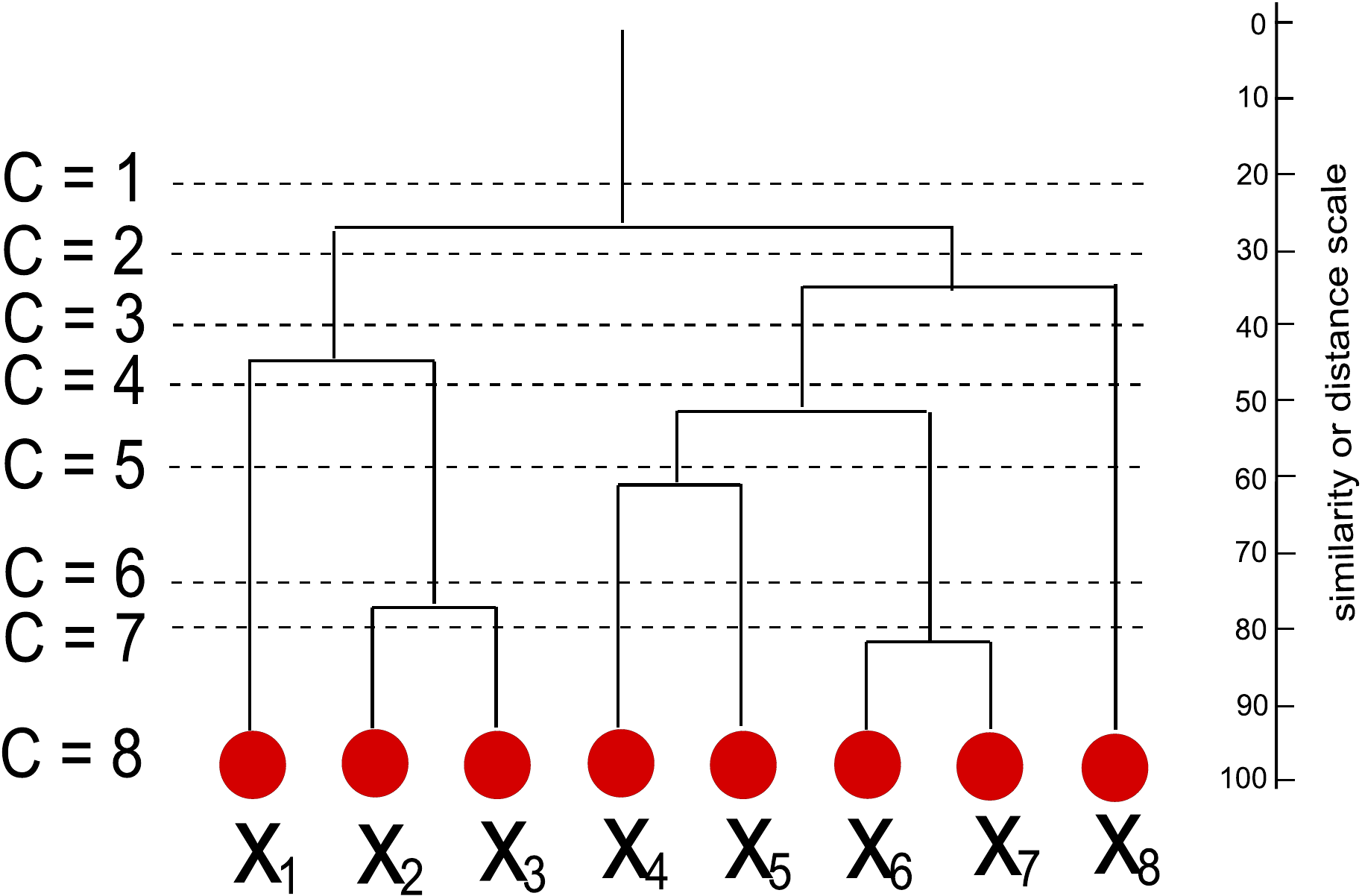}
  \caption{A dendrogram for a simple situation involving eight objects, adapted from \cite{DUDA2001}.}
\label{fig:dendrogram}
\end{figure}

To show how the objects are grouped, hierarchical clustering can be
represented by a corresponding tree, called \textit{dendrogram}. Figure
\ref{fig:dendrogram} illustrates a dendrogram representing the results
of hierarchical clustering for a problem with eight objects. The
measure of similarity among clusters can be observed in the vertical
axis. The different number of classes can be obtained by horizontally
cutting the dendrogram at different values of similarity or distance. 

Hence, to perform hierarchical cluster analysis it is necessary to
define three main parameters. The first regards how to quantify the
similarity between every pair of objects in the data set, that is, how
to calculate the distance between the objects. Euclidean distance,
which is frequently used, will be adopted in this work, but other
possible distances are cityblock, cheesboard, mahalanobis and so
on. The second parameter is the linkage method, which establishes how
to measure the distance between two sets. The linkage method can be
used to link pairs of objects that are similar and then to form the
hierarchical cluster tree. There are many possibilities for doing so,
some of the most popular are: single linkage, complete linkage, group
linkage, centroid linkage, mean linkage and ward's linkage
(\cite{JAIN1988, ANDERBERG1973, ROMESBURG1990, COSTA2001}). Ward's linkage uses the intraclass dispersion as a
clustering criterion. Pairs of objects are merged in such a way to
guarantee the smallest increase in the intraclass dispersion. This
clustering approach has been sometimes identified as corresponding to
the best hierarchical method \cite{KUIPER1975, BLASHFILED1976, MOJENA1975} and will be used in this
work. Actually, it is particularly interesting to analyse the
intraclass dispersion in an unsupervised classification procedure in
order to identify common and different characteristics when compared
to the supervised classification. The third parameter concerns the
number of desired clusters, an issue which is directly related to
where to cut the dendrogram into clusters, as illustrated by $C$ in
Figure \ref{fig:dendrogram}.

\clearpage

\section{The Kappa Coefficient} \label{sec:AKappa}

\begin{table*}
    \centering
    \caption{Classification performance according to kappa}
    \footnotesize
        \begin{tabular}{l c c}
        \hline
        \textbf{Kappa} & \textbf{Classification performance} \\
        \hline\hline
        $\hat{k}   \le 0$           & Poor \\ 
        $0 < \hat{k}    \le 0.2$     & Slight \\
        $0.2 < \hat{k}    \le 0.4$   & Fair \\
        $0.4 < \hat{k}    \le 0.6$   & Moderate \\
        $0.6 < \hat{k}    \le 0.8$   & Substantial \\
        $0.8 < \hat{k}    \le 1.0$   & Almost Perfect \\
          \hline\hline
        \end{tabular}
        \label{tab:Kappa}
\end{table*}

The kappa coefficient was first proposed by Cohen \cite{COHEN1960}. In
the context of supervised classification, this coefficient determines
the degree of agreement a \textit{posteriori}. This means that it
quantifies the agreement between objects previously known (ground
truth) and the result obtained by the classifier. The better the
classification accuracy, the higher the degree of concordance, and,
consequently, the higher the value of kappa. The kappa coefficient is
computed from the confusion matrix as follows \cite{CONGALTON1991}:

\begin{equation}
\hat{k}  = \frac{{N\sum\nolimits_{i = 1}^C {c_{ii}  - \sum\nolimits_{i = 1}^C {x_{i + } x_{ + i} } } }}{{N^2  - \sum\nolimits_{i = 1}^C {x_{i + } x_{ + i} } }}
\end{equation}

where ${x_{i + } }$ is the sum of elements from line $i$, ${x_{ + i}
}$ is the sum of elements from column $i$, $C$ is the number of
classes (confusion matrix is $C$ x $C$), and $N$ is the total number
of objects. The kappa variance can be calculated as:

\begin{equation}
\hat{\sigma}_k^2  = \frac{1}{N}\left[ {\frac{{\theta _1 \left( {1 - \theta _1 } \right)}}{{\left( {1 - \theta _2 } \right)^2 }} + \frac{{2\left( {1 - \theta _1 } \right)\left( {2\theta _1 \theta _2  - \theta _3 } \right)}}{{\left( {1 - \theta _2 } \right)^3 }} + \frac{{\left( {1 - \theta _1 } \right)^2 \left( {\theta _4  - 4\theta _2^2 } \right)}}{{\left( {1 - \theta _2 } \right)^4 }}} \right]
\end{equation}
 
where 

\begin{equation}
\eqalign{
\theta_{1} = \frac{1}{N}\sum_{i=1}^{C}x_{ii} \cr
\theta_{2} = \frac{1}{N^{2}}\sum_{i=1}^{C}x_{i+}x_{+i} \cr
\theta_{3} = \frac{1}{N^{2}}\sum_{i=1}^{C}x_{ii}\left( x_{i+} + x_{+i} \right) \cr
\theta_{4} = \frac{1}{N^{3}}\sum_{i=1}^{C}\sum_{j=1}^{C}x_{ij}\left( x_{j+} + x_{+i} \right)^{2}}
\end{equation}
 
This statistics indicates that, when $\hat{k} \le 0$
there is not any agreement, and when $\hat{k} = 1$
the agreement is total. Some authors suggest interpretations according
to the value obtained by the coefficient kappa. Table
\ref{tab:Kappa} shows one possible interpretation, proposed by
\cite{LANDIS1977}.

\clearpage
\section*{References}
\bibliographystyle{unsrt}
\bibliography{References}

\end{document}